%% file: Dp_to_Klenu.tex
\documentclass[aps,prd,twocolumn,showpacs,floatfix,byrevtex]{revtex4-1}
\usepackage{times}
\usepackage{amsmath}
\usepackage{graphicx}
\usepackage{color}
\usepackage{lineno}
\usepackage{bm}
\usepackage[colorlinks=true,linkcolor=blue,citecolor=blue]{hyperref}

\begin{document}
\normalsize
\hyphenpenalty=5000
\tolerance=1000


\def\eref#1{(\ref{#1})}
\def\Journal#1#2#3#4{#1 {\bf #2}, #3 (#4)}
\def\IJMP{Int. J. Mod. Phys. A}
\def\NCA{Nuovo Cimento}
\def\NIM{Nucl. Instrum. Meth.}
\def\NIMA{Nucl. Instrum. Meth. A}
\def\NPB{Nucl. Phys. B}
\def\PLB{Phys. Lett. B}
\def\PRL{Phys. Rev. Lett.}
\def\PRD{Phys. Rev. D}
\def\PRP{Phys. Rep.}
\def\ZPC{Z. Phys. C}
\def\EPJC{Eur. Phys. J. C}
\def\HEPNP{HEP \& NP}
\def\CPC{Chin. Phys. C}
\def\RN#1{\uppercase\expandafter{\romannumeral#1}}
\def\TeV{\hbox{$\,\hbox{TeV}$}}
\def\GeV{\hbox{$\,\hbox{GeV}$}}
\def\gev{\hbox{$\,\hbox{GeV/$c^2$}$}}
\def\MeV{\hbox{$\,\hbox{MeV}$}}
\def\mev{\hbox{$\,\hbox{MeV/$c^2$}$}}
\def\keV{\hbox{$\,\hbox{keV}$}}
\def\eV{\hbox{$\,\hbox{eV}$}}
\def\nb{\hbox{$\,\hbox{nb}$}}
\def\pb{\hbox{$\,\hbox{pb}$}}
\def\fb{\hbox{$\,\hbox{fb}$}}
\def\m{\hbox{$\,\hbox{m}$}}
\def\mum{\hbox{$\,\hbox{$\mu$m}$}}
\def\cm{\hbox{$\,\hbox{cm}$}}
\def\mm{\hbox{$\,\hbox{mm}$}}
\def\sec{\hbox{$\,\hbox{sec}$}}
\def\s{\hbox{$\,\hbox{s}$}}
\def\ps{\hbox{$\,\hbox{ps}$}}
\def\mus{\hbox{$\,\hbox{$\mu$s}$}}
\def\ns{\hbox{$\,\hbox{ns}$}}

\newcommand{\jpsi}{J/\psi}
\newcommand{\psip}{\psi^{\prime}}
\newcommand{\psipp}{\psi(3770)}
\newcommand{\ks}{K^{0}_{S}}
\newcommand{\kl}{K^{0}_{L}}
\newcommand{\kz}{K^{0}}
\newcommand{\kzb}{\bar K^{0}}
\newcommand{\dz}{D^{0}}
\newcommand{\dzb}{\bar D^{0}}
\newcommand{\ddb}{D \bar D}
\newcommand{\ee}{e^{+}e^{-}}
\newcommand{\pip}{\pi^{+}}
\newcommand{\pim}{\pi^{-}}
\newcommand{\piz}{\pi^{0}}
\newcommand{\mumu}{\mu^{+}\mu^{-}}
\newcommand{\pipi}{\pi^{+}\pi^{-}}
\newcommand{\kk}{K^{+}K^{-}}
\newcommand{\gam}{\gamma}
\newcommand{\eff}{\epsilon}
\newcommand{\mbc}{M_{\mathrm{BC}}}
\newcommand{\beq}{\begin{equation}}
\newcommand{\eeq}{\end{equation}}
\newcommand{\bitm}{\begin{itemize}}
\newcommand{\eitm}{\end{itemize}}


\title{\boldmath Study of decay dynamics and $CP$ asymmetry in $D^+ \to \kl e^+ \nu_e$ decay}

\input{authors_sep2015}

\date{\today}

\begin{abstract}

  Using $2.92$\fb$^{-1}$ of electron-positron annihilation data
  collected at $\sqrt{s} = 3.773\GeV$ with the BESIII detector, we
  obtain the first measurements of the absolute branching fraction
  $\mathcal{B}(D^+ \to \kl e^+ \nu_e) = (4.481 \pm
  0.027(\mathrm{stat.})  \pm 0.103(\mathrm{sys.}))\%$ and the $CP$
  asymmetry $A_{CP}^{D^+ \to \kl e^+ \nu_e} = (-0.59 \pm
  0.60(\mathrm{stat.}) \pm 1.48(\mathrm{sys.}))\%$. From the $D^+ \to
  \kl e^+ \nu_e$ differential decay rate distribution, the product of
  the hadronic form factor and the magnitude of the CKM matrix element,
  $f_{+}^{K}(0)|V_{cs}|$, is determined to be $0.728 \pm
  0.006(\mathrm{stat.}) \pm 0.011(\mathrm{sys.})$. Using $|V_{cs}|$
  from the SM constrained fit with the measured
  $f_{+}^{K}(0)|V_{cs}|$, $f_{+}^{K}(0) = 0.748 \pm
  0.007(\mathrm{stat.}) \pm 0.012(\mathrm{sys.})$ is obtained, and
  utilizing the unquenched LQCD calculation for $f_{+}^{K}(0)$,
  $|V_{cs}| = 0.975 \pm 0.008(\mathrm{stat.}) \pm 0.015(\mathrm{sys.})
  \pm 0.025(\mathrm{LQCD})$.

\end{abstract}


\maketitle



\section{Introduction}
\label{sec:intro}

In the Standard Model (SM), violation of the combined
charge-conjugation and parity symmetries ($CP$) arises from a
nonvanishing irreducible phase in the Cabibbo-Kobayashi-Maskawa (CKM)
flavor-mixing matrix~\cite{ckm, cabibbo}. Although in the SM, $CP$ violation in
the charm sector is expected to be very small, $\mathcal{O}(10^{-3})$
or below~\cite{cpv}, reference~\cite{xingzz} finds that $K^0$-$\bar
K^0$ mixing will give rise to a clean $CP$ violation signal of
magnitude of $-2\mathrm{Re}(\epsilon) \approx -3.3 \times 10^{-3}$ in
the semileptonic decays $D^+ \to \kl(\ks) e^+ \nu_e$.

Semileptonic decays of mesons allow determination of various important
SM parameters, including elements of the CKM matrix, which in turn
allows the physics of the SM to be tested at its most fundamental
level. In the limit of zero electron mass, the differential decay rate
for a $D$ semileptonic decay with a pseudoscalar meson $P$ is given by
\begin{equation}
    \label{eqn:dgdq2}
    \frac{d\Gamma(D \to P e \nu_e)}{dq^2} =
    \frac{G_F^2|V_{cs(d)}|^2}{24\pi^3} p^3 |f_+(q^2)|^2,
\end{equation}
where $G_F$ is the Fermi constant, $V_{cs(d)}$ is the relevant CKM
matrix element, $p$ is the momentum of the daughter meson in the rest
frame of the parent $D$, $f_+(q^2)$ is the form factor, and $q^2$ is
the invariant mass squared of the lepton-neutrino system.

In this paper, the first measurements of the absolute branching
fraction and the $CP$ asymmetry for the decay $D^+ \to \kl e^+ \nu_e$, as well as the form-factor parameters for three
different theoretical models that describe the weak hadronic charged
currents in $D^+ \to \kl e^+ \nu_e$ are presented. The paper is organized as
follows: The BESIII detector and data sample are described in
Sec.~\ref{sec:sample}. The analysis technique is introduced in
Sec.~\ref{sec:reconstruction}. In Secs.~\ref{sec:brfraction}
and~\ref{sec:formfactor} the measurements of the absolute branching
fraction, the $CP$ asymmetry and the form-factor parameters for the
decay $D^+ \to \kl e^+ \nu_e$ are described. Finally, a summary is
provided in Sec.~\ref{sec:summary}.

\section{The BESIII Detector and Data Sample}
\label{sec:sample}

The analysis presented in this paper is based on a data sample with an
integrated luminosity of 2.92\fb$^{-1}$~\cite{luminosity} collected
with the BESIII detector~\cite{detector} at the center-of-mass energy
of $\sqrt{s}=3.773\GeV$.  The BESIII detector is a general-purpose
detector at the BEPCII~\cite{accelerator} double storage
rings. The detector has a geometrical acceptance of 93\% of the full
solid angle.  We briefly describe the components of BESIII from the
interaction point (IP) outwards.  A small-cell multilayer drift
chamber (MDC), using a helium-based gas to measure momenta and
specific ionization of charged particles, is surrounded by a
time-of-flight (TOF) system based on plastic scintillators which
determines the time of flight of charged particles.  A CsI(Tl)
electromagnetic calorimeter (EMC) detects electromagnetic
showers. These components are all situated inside a superconducting
solenoid magnet, which provides a 1.0\,T magnetic field parallel to
the beam direction.  Finally, a multilayer resistive plate counter
system installed in the iron flux return yoke of the magnet is used to
track muons.  The momentum resolution for charged tracks in the MDC is
0.5\% for a transverse momentum of 1\,GeV/$c$. The energy resolution for
showers in the EMC is 2.5\% for 1\,GeV photons. More details on the
features and capabilities of BESIII can be found
elsewhere~\cite{detector}.

The performance of the BESIII detector is simulated using a {\sc
  geant4}-based~\cite{geant4} Monte Carlo (MC) program. To develop
selection criteria and test the analysis technique, several MC samples
are used. For the production of $\psi(3770)$, the {\sc
  kkmc}~\cite{kkmc} package is used; the beam energy spread and the
effects of initial-state radiation (ISR) are included.
Final-state radiation (FSR) of charged tracks is
  taken into account with the {\sc photos} package~\cite{photos}.
$\psipp \to \ddb$ events are generated using {\sc
  evtgen}~\cite{eventgen, beseventgen}, and each $D$ meson is allowed
to decay according to the branching fractions in the Particle Data
Group (PDG)~\cite{pdg}. We refer to this as the ``generic MC.'' The
equivalent luminosity of the MC samples is about 10
times that of the data. A sample of $\psipp \to \ddb$ events, in which
the $D$ meson decays to the signal semileptonic mode and the $\bar D$
decays to one of the hadronic final states used in the tag
reconstruction, is referred to as the ``signal MC''. In both the
generic and signal MC samples, the semileptonic decays are generated
using the modified pole parametrization~\cite{becirevic} (see
Sec.~\ref{subsec:ff para}).

\section{Event Selection}
\label{sec:reconstruction}

At the $\psi(3770)$ peak, $\ddb$ pairs are produced. First, we select
the single-tag (ST) sample in which a $D^-$ is reconstructed in a
hadronic decay mode. From the ST sample, the double-tag (DT) events of
$D^+ \to \kl e^+ \nu_e$ are selected. The numbers of the ST and DT
events are given by
\begin{equation}
    \begin{split}
    N_{\mathrm{ST}} &= N_{D^{+}D^{-}} \mathcal{B}_{\mathrm{tag}} \eff_{\mathrm{ST}}, \\
    N_{\mathrm{DT}} &= N_{D^{+}D^{-}} \mathcal{B}_{\mathrm{tag}} \mathcal{B}_{\mathrm{sig}} \eff_{\mathrm{DT}},
    \end{split}
    \label{eqn:br}
\end{equation}
where $N_{D^{+}D^{-}}$ is the number of $D^{+}D^{-}$ pairs produced,
$N_{\mathrm{ST}}$ and $N_{\mathrm{DT}}$ are the numbers of the ST and
DT events, $\eff_{\mathrm{ST}}$ and $\eff_{\mathrm{DT}}$ are the
corresponding efficiencies, and $\mathcal{B}_{\mathrm{tag}}$ and
$\mathcal{B}_{\mathrm{sig}}$ are the branching fractions of the
hadronic tag decay and the signal decay. In this analysis, the
charge-dependent branching fractions are measured, so there is no
factor of two in Eq.~(\ref{eqn:br}). From Eq.~(\ref{eqn:br}), we
obtain
\begin{equation}
    \mathcal{B}_{\mathrm{sig}} = \frac{N_{\mathrm{DT}} /
    \eff_{\mathrm{DT}}}{N_{\mathrm{ST}} / \eff_{\mathrm{ST}}} =
    \frac{N_{\mathrm{DT}} / \eff}{N_{\mathrm{ST}}},
\end{equation}
where $\eff = \eff_{\mathrm{DT}} / \eff_{\mathrm{ST}}$ is the
efficiency of finding a signal candidate in the presence of a ST $D$,
which is obtained from generic MC simulations.

\subsection{Selection of ST events}
\label{subsec:selection}
Each charged track is required to satisfy $|\cos\theta|<0.93$, where
$\theta$ is the polar angle with respect to the beam axis. Charged
tracks other than those from the $\ks$ are required to have their
points of closest approach to the beamline within $10 \cm$ from the
IP along the beam axis and within $1 \cm$ in the plane perpendicular
to the beam axis.  Particle identification for charged hadrons $h$ ($h
= \pi, K$) is accomplished by combining the measured energy loss
($dE/dx$) in the MDC and the flight time obtained from the TOF to form
a likelihood $\mathcal{L}$($h$) for each hadron hypothesis.  The
$K^\pm$ ($\pi^\pm$) candidates are required to satisfy
$\mathcal{L}(K)>\mathcal{L}(\pi)$ ($\mathcal{L}(\pi)>
\mathcal{L}(K)$).

The $\ks$ candidates are selected from pairs of oppositely charged
tracks which satisfy a vertex-constrained fit to a common vertex.
The vertices are required to be within $20 \cm$ of the IP along the
beam direction; no constraint in the transverse plane is
applied. Particle identification is not required, and the two charged
tracks are assumed to be pions.  We require $|M_{\pi^+\pi^-} -
M_{\ks}| < 12 \MeV/c^2$, where $M_{\ks}$ is the nominal $\ks$
mass~\cite{pdg} and $12 \MeV/c^2$ is about 3 standard deviations of
the observed $\ks$ mass resolution. Lastly, the $\ks$ candidate must
have a decay length more than 2 standard deviations of the vertex
resolution away from the IP.

Reconstructed EMC showers that are separated from the extrapolated
positions of any charged tracks by more than $10^{\circ}$ are taken as
photon candidates. The energy deposited in the nearby TOF counters is
included to improve the reconstruction efficiency and energy
resolution. Photon candidates must have a minimum energy of $25 \MeV$
for barrel showers ($|\cos\theta| < 0.80$) and $50 \MeV$ for end-cap
showers ($0.86 < |\cos\theta| < 0.92$). The shower timing is required
to be no later than $700 \ns$ after the reconstructed event start time
to suppress electronic noise and energy deposits unrelated to the
event.

The $\pi^0$ candidates are reconstructed from pairs of photons, and
the invariant mass $M_{\gamma\gamma}$ is required to satisfy $0.110 <
M_{\gamma \gamma} < 0.155 \GeV/c^2$. The invariant mass of two photons
is constrained to the nominal $\piz$ mass~\cite{pdg} by a kinematic
fit, and the $\chi^2$ of the kinematic fit is required to be less than
20.

We form $D^{\pm}$ candidates decaying into final hadronic states of
$K^{\mp} \pi^{\pm} \pi^{\pm}$, $K^{\mp} \pi^{\pm} \pi^{\pm} \piz$,
$\ks \pi^{\pm} \piz$, $\ks \pi^{\pm} \pi^{\pm} \pi^{\mp}$, $\ks
\pi^{\pm}$, and $K^+ K^- \pi^{\pm}$. Two variables are used to
identify valid ST $D$ candidates: $\Delta E \equiv E_D -
E_{\mathrm{beam}}$, the energy difference between the energy of the ST
$D$ ($E_D$) and the beam energy ($E_{\mathrm{beam}}$), and the
beam-constrained mass $\mbc \equiv \sqrt{E^{2}_{\mathrm{beam}}/c^4 -
  |\vec{p}_{D}|^2/c^{2}}$, where $\vec{p}_{D}$ is the momentum of the
$D$. The ST $D$ signal should peak at the nominal $D$ mass in the $\mbc$
distribution and around zero in the $\Delta E$ distribution.  We only
accept one candidate per mode; when multiple candidates are present in
an event, the one with the smallest $|\Delta E|$ is kept. Backgrounds
are suppressed by the mode-dependent $\Delta E$ requirements listed in
Table~\ref{table:st-de}.

\begin{table}[htbp]
    \centering
    \caption{\label{table:st-de}Requirements on $\Delta E$ for the ST
    $D$ candidates. The limits are set at approximately 3 standard
    deviations of the $\Delta E$ resolution.}
    \begin{tabular}{lc}\hline\hline
        \multicolumn{1}{c}{Mode} &
        \multicolumn{1}{c}{Requirement (GeV)} \\
        \hline
        $D^{\pm} \to K^{\mp} \pi^{\pm} \pi^{\pm}$             & $-0.030 < \Delta E < 0.030$\\
        $D^{\pm} \to K^{\mp} \pi^{\pm} \pi^{\pm} \piz$        & $-0.052 < \Delta E < 0.039$\\
        $D^{\pm} \to \ks \pi^{\pm} \piz$                      & $-0.057 < \Delta E < 0.040$\\
        $D^{\pm} \to \ks \pi^{\pm} \pi^{\pm} \pi^{\mp}$       & $-0.034 < \Delta E < 0.034$\\
        $D^{\pm} \to \ks \pi^{\pm}$                           & $-0.032 < \Delta E < 0.032$\\
        $D^{\pm} \to K^+ K^- \pi^{\pm}$                       & $-0.030 < \Delta E < 0.030$\\
        \hline\hline
    \end{tabular}
\end{table}

The ST yields of data are determined by binned maximum likelihood
fits to the $\mbc$ distributions. The signal MC line shape is used to
describe the $D$ signal, and an ARGUS~\cite{argus} function is used to model
the combinatorial backgrounds from the continuum light hadron
production, $\gamma_{\mathrm{ISR}}\psi(3686)$,
$\gamma_{\mathrm{ISR}}J/\psi$ and non-signal $\ddb$ decays. A Gaussian
function, with the standard deviation and the central value as free
parameters, is convoluted with the line shape to account for imperfect
modeling of the detector resolution and beam energy.

\begin{figure*}[htbp]
	\centering
    \includegraphics[height=3.5cm]{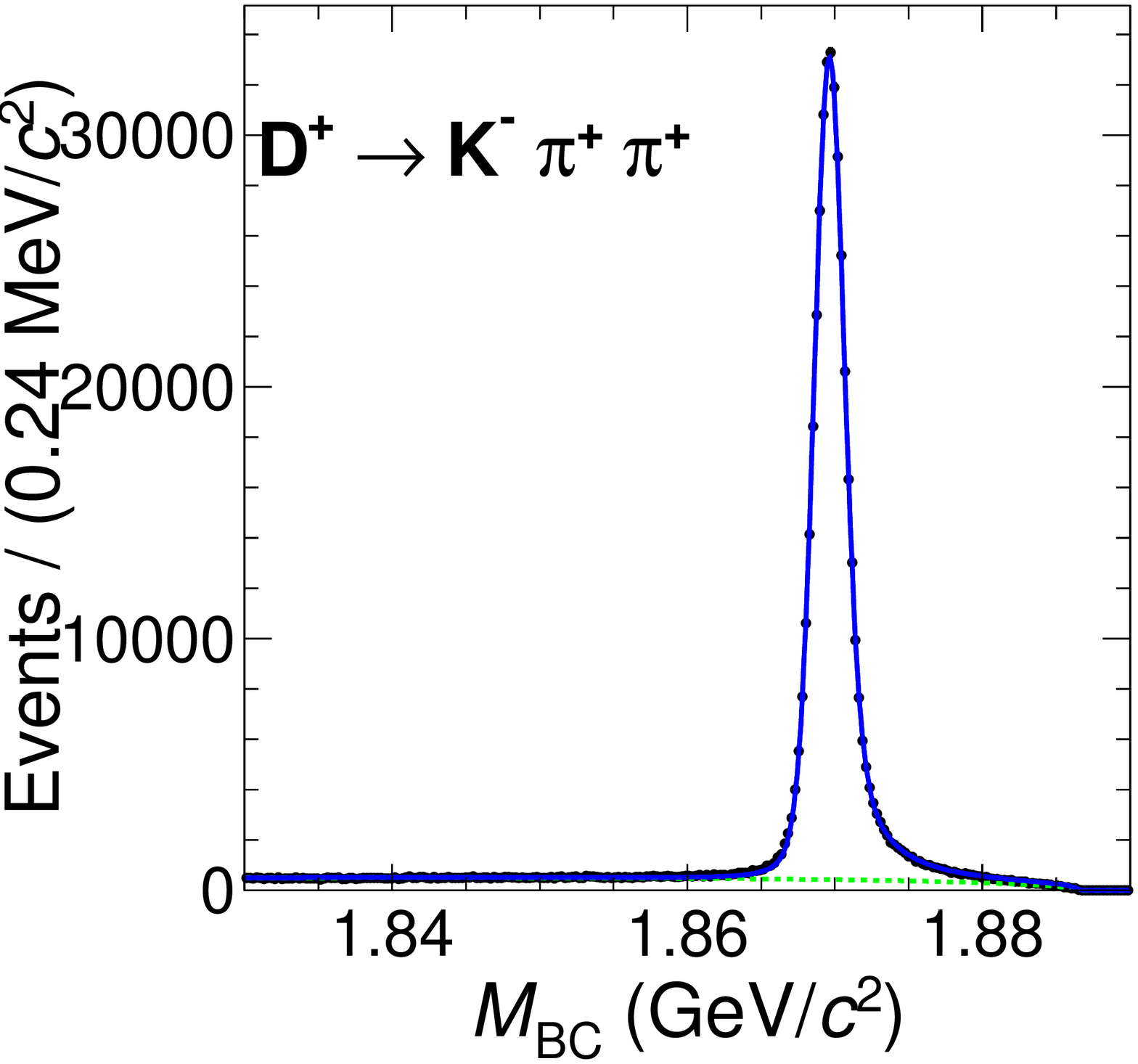}
    \includegraphics[height=3.5cm]{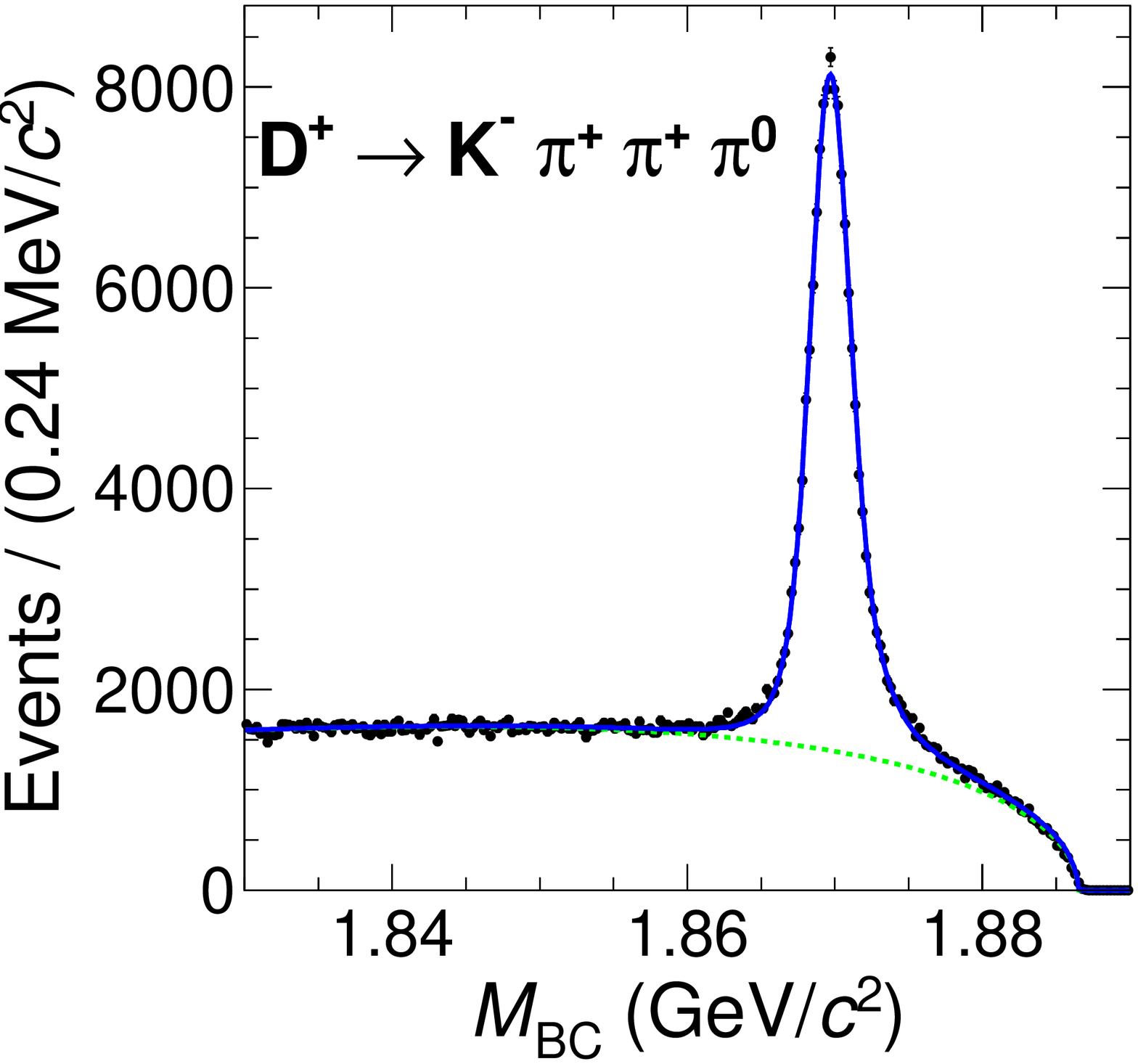}
    \includegraphics[height=3.5cm]{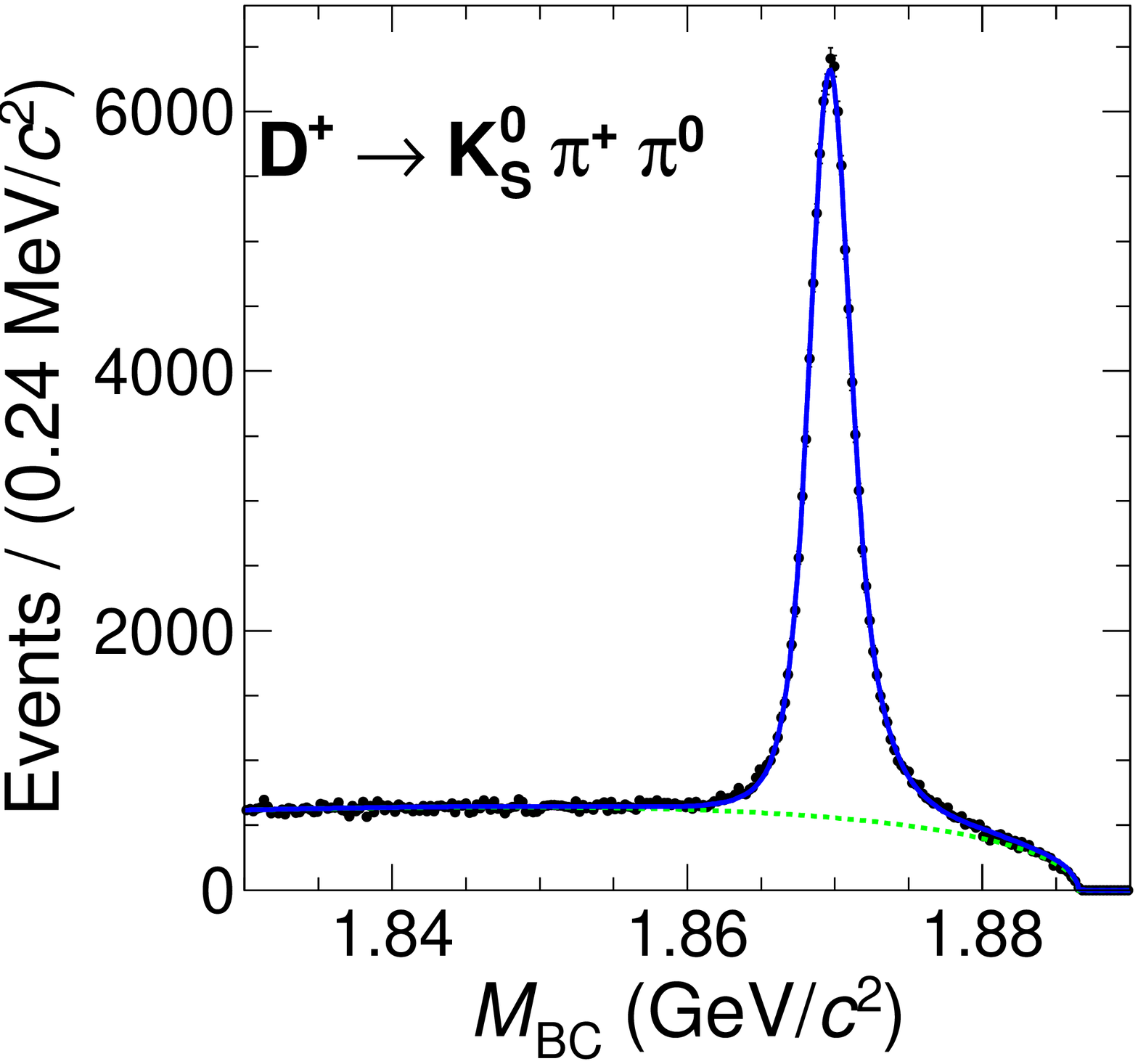}\\
    \includegraphics[height=3.5cm]{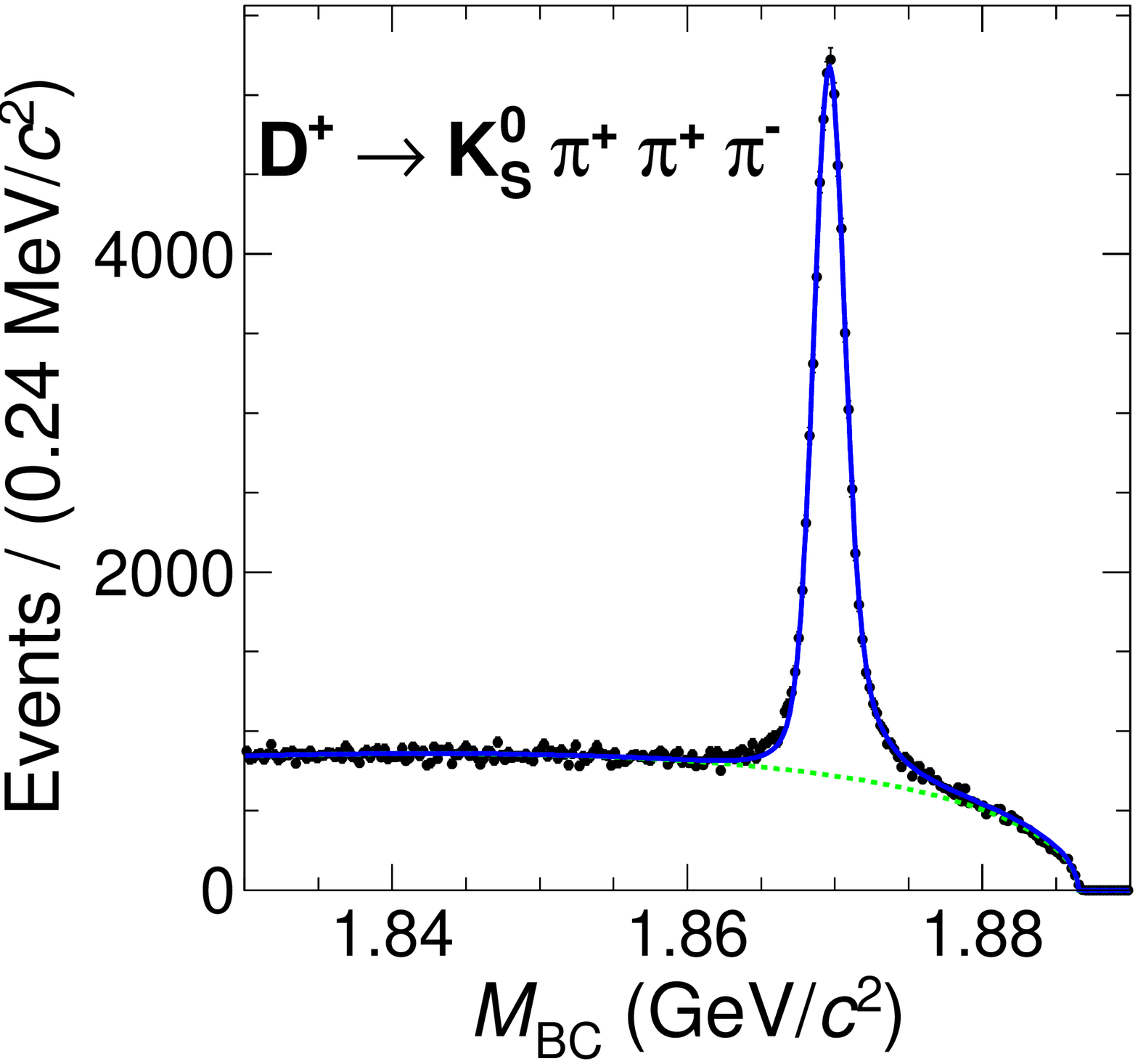}
    \includegraphics[height=3.5cm]{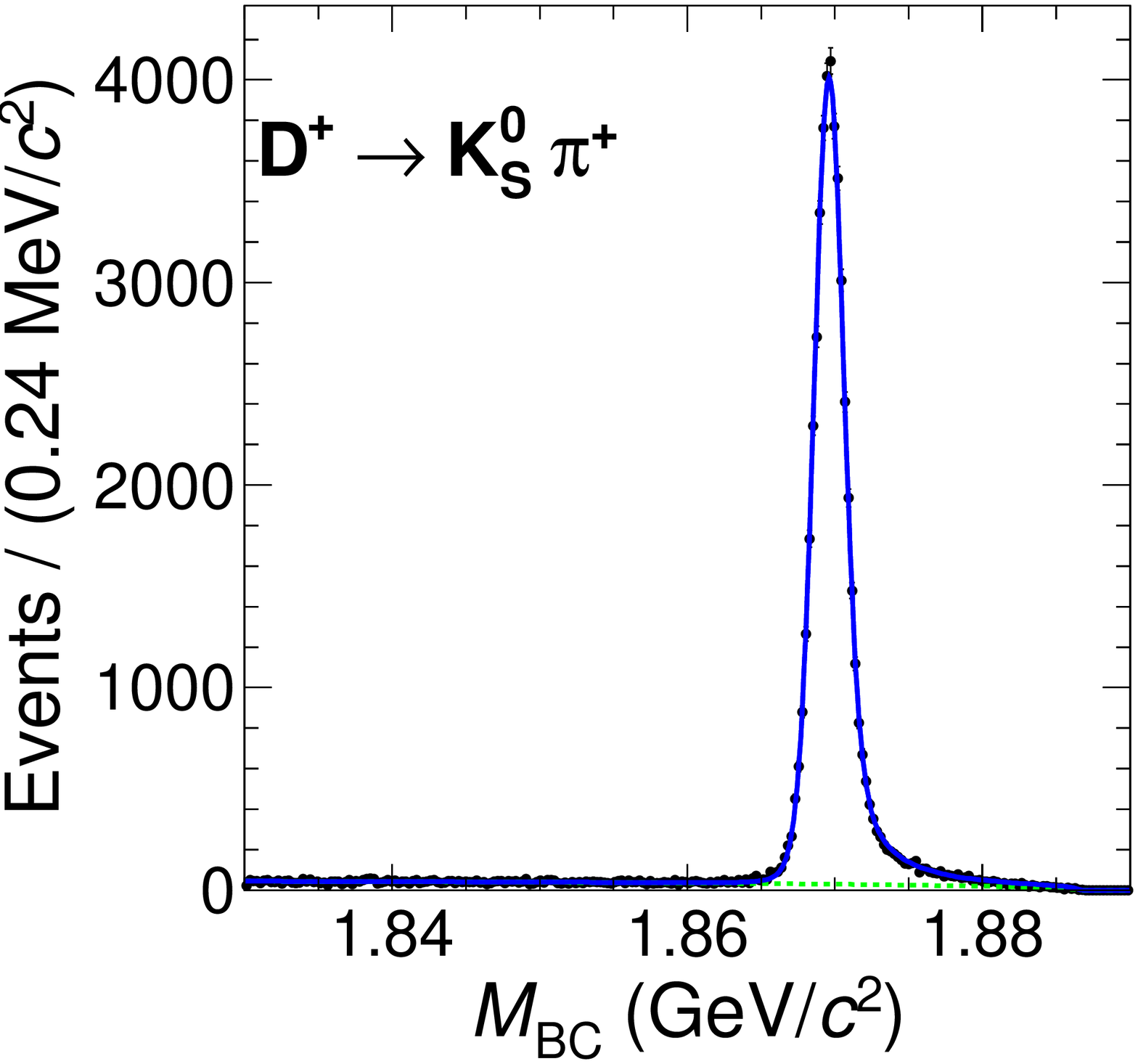}
    \includegraphics[height=3.5cm]{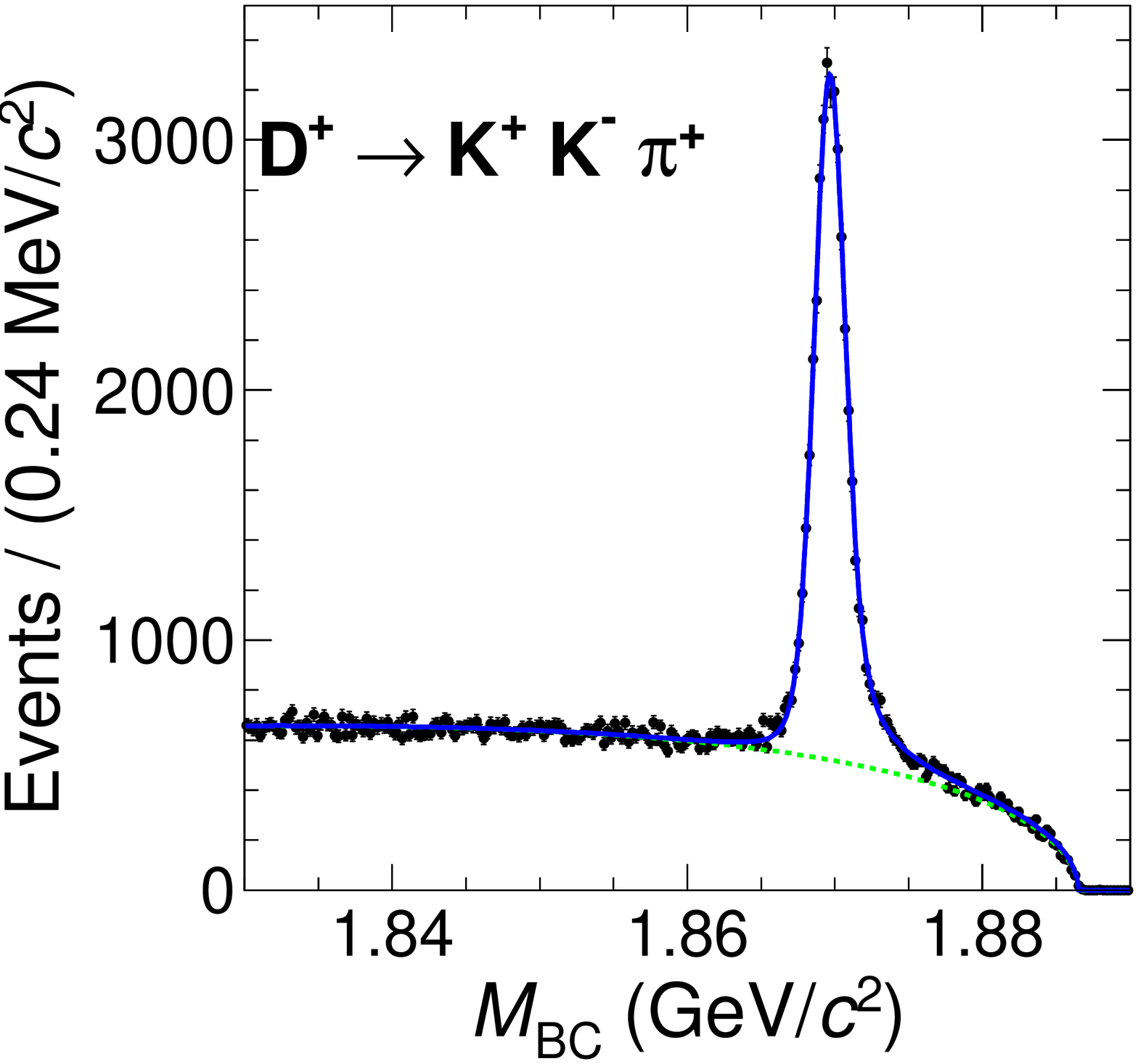}
	\caption{\label{fig:st-mbc-data-dp}Fits to the $\mbc$
	distributions of the ST $D^+$ candidates for data. The dots
	with error bars are for data, and the blue solid curves are the
	results of the fits. The green dashed curves are the fitted
	backgrounds.}
\end{figure*}

\begin{figure*}[htbp]
	\centering \includegraphics[height=3.5cm]{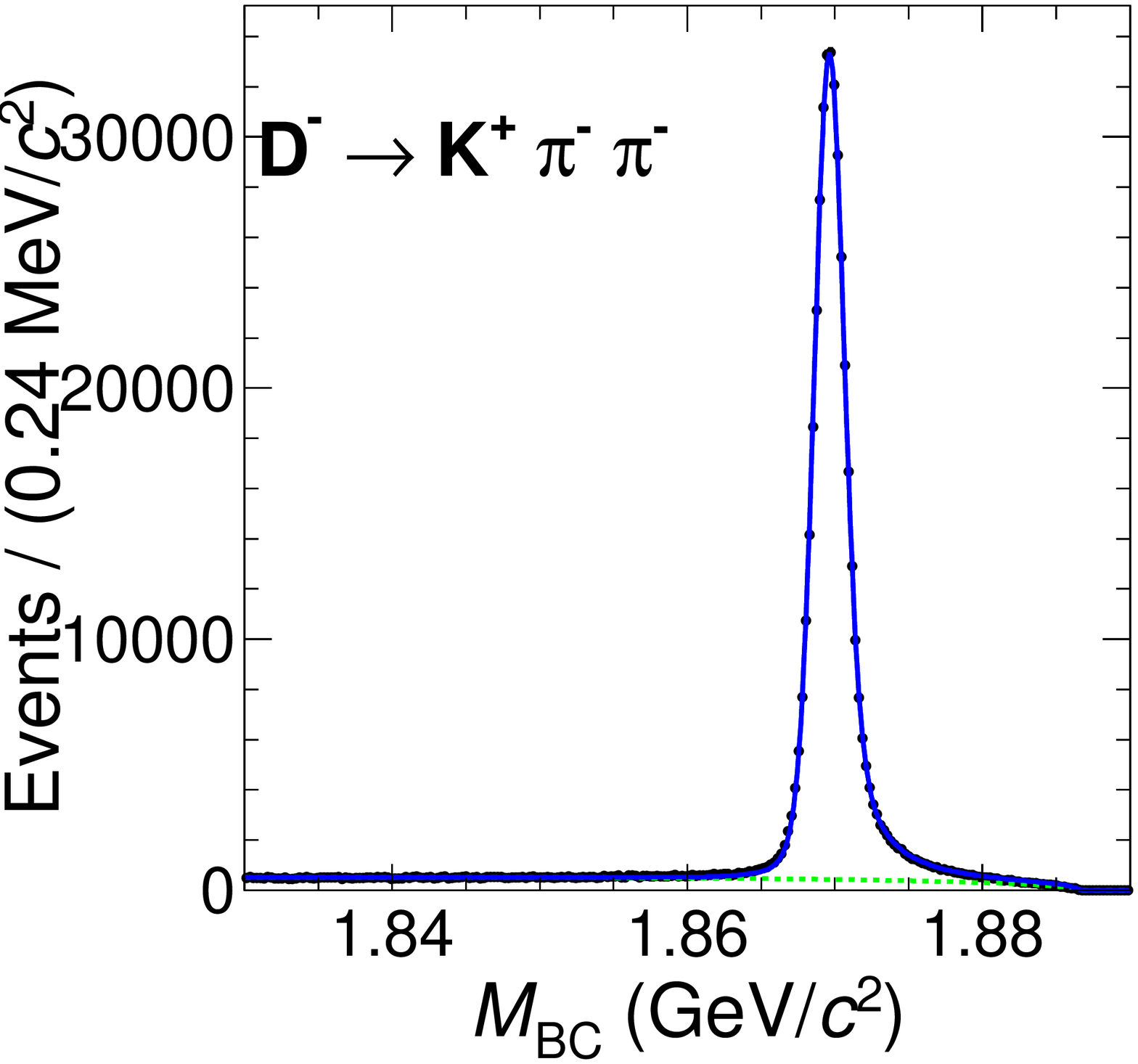}
    \includegraphics[height=3.5cm]{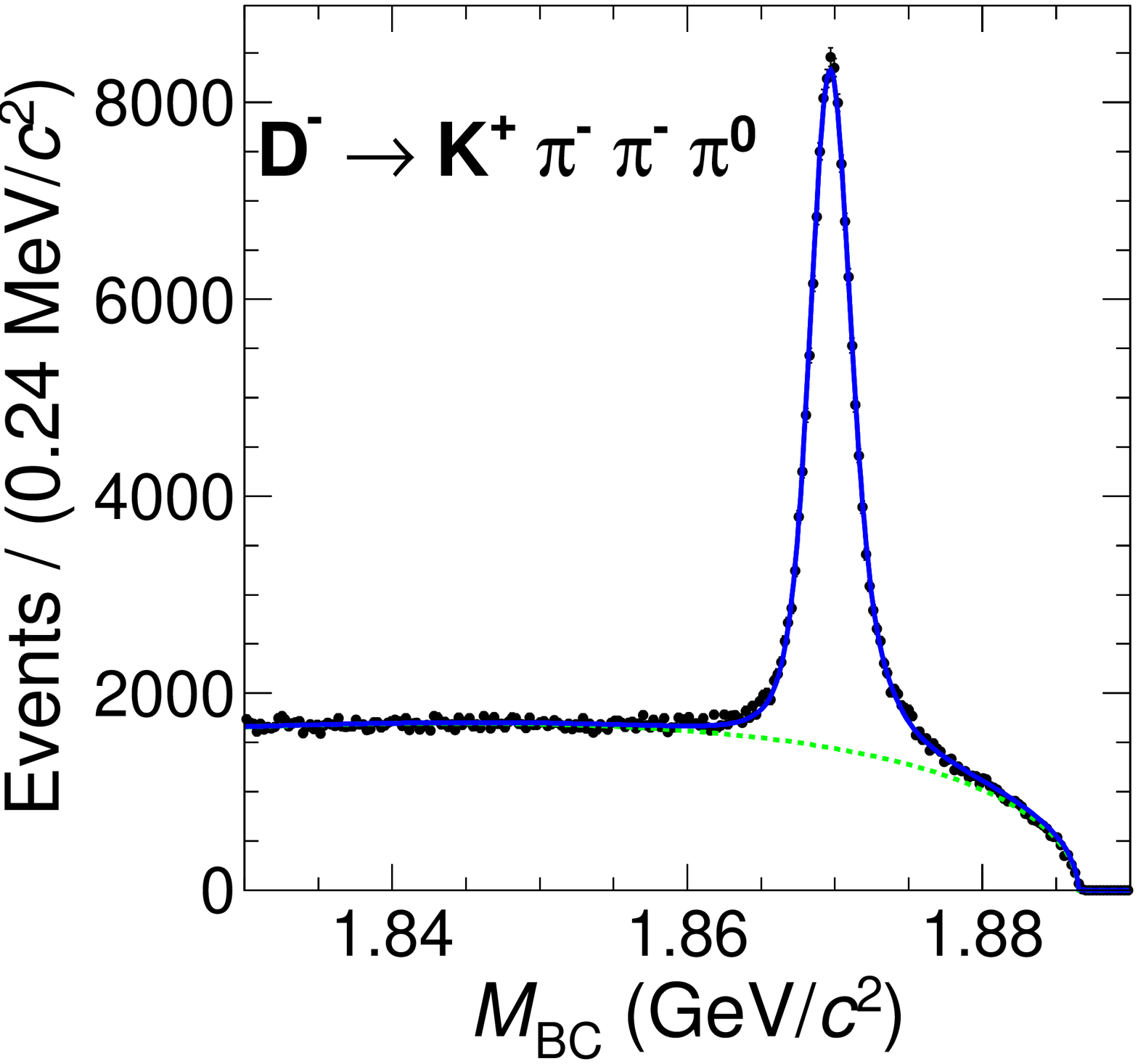}
    \includegraphics[height=3.5cm]{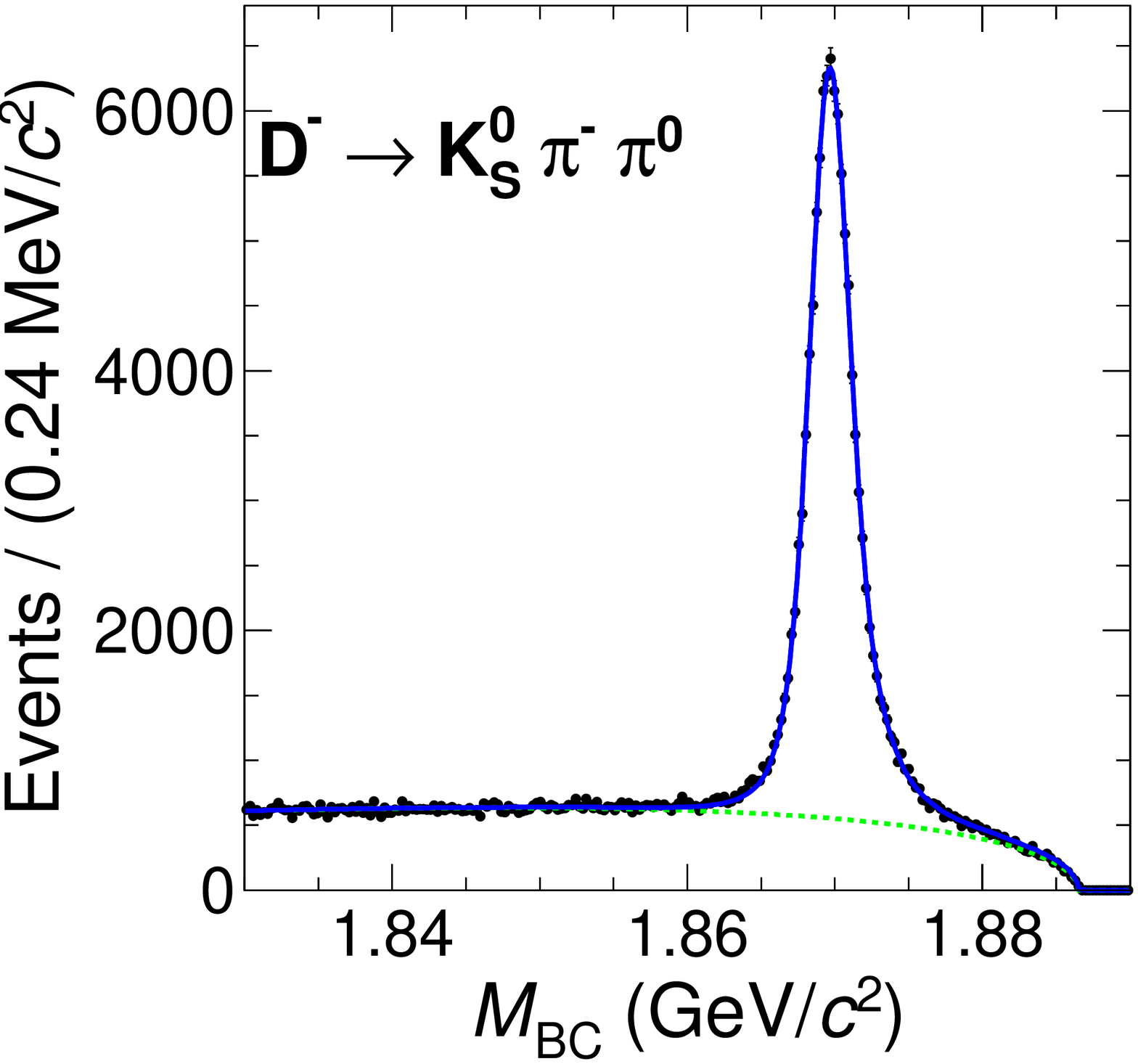}\\
    \includegraphics[height=3.5cm]{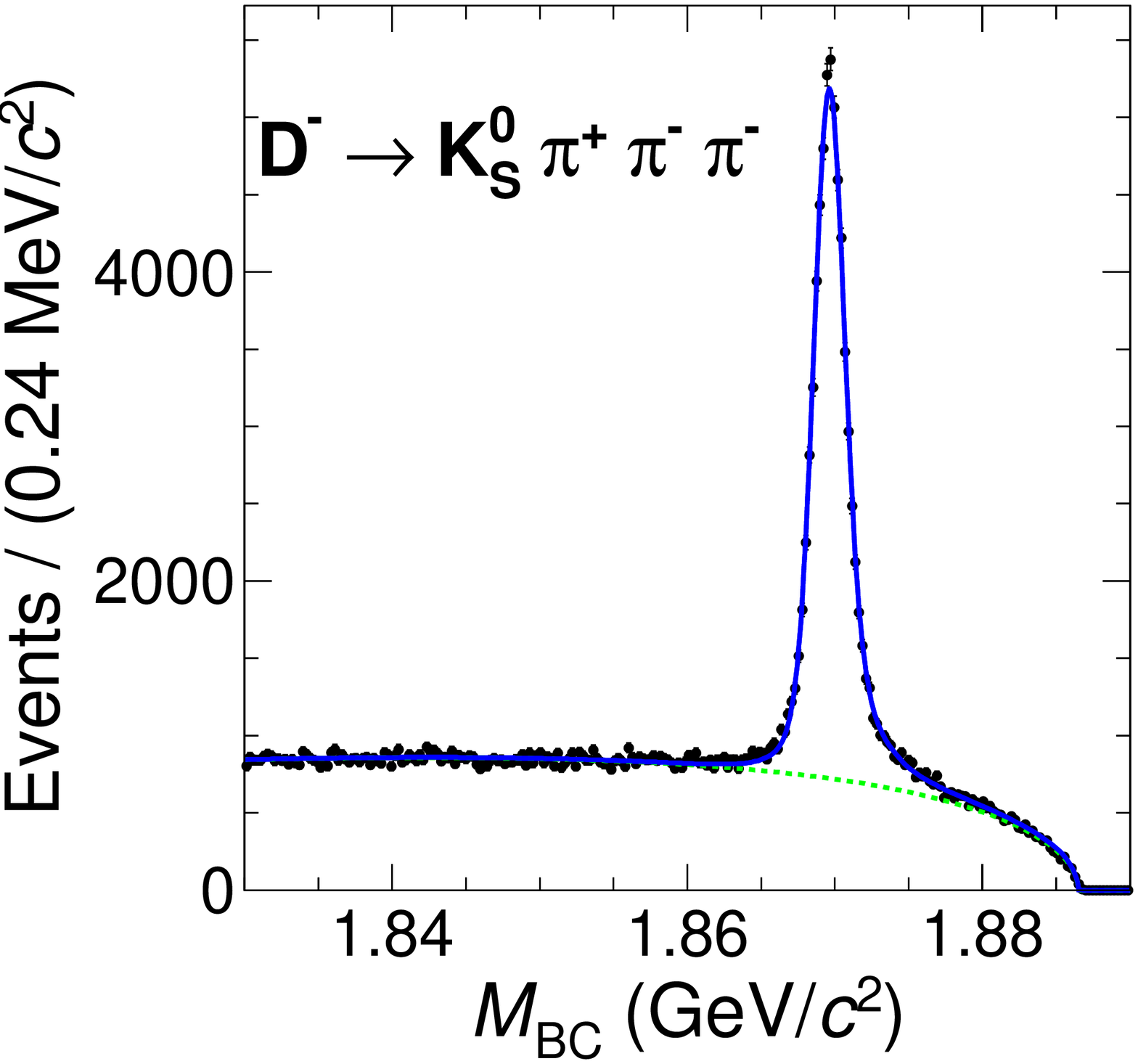}
    \includegraphics[height=3.5cm]{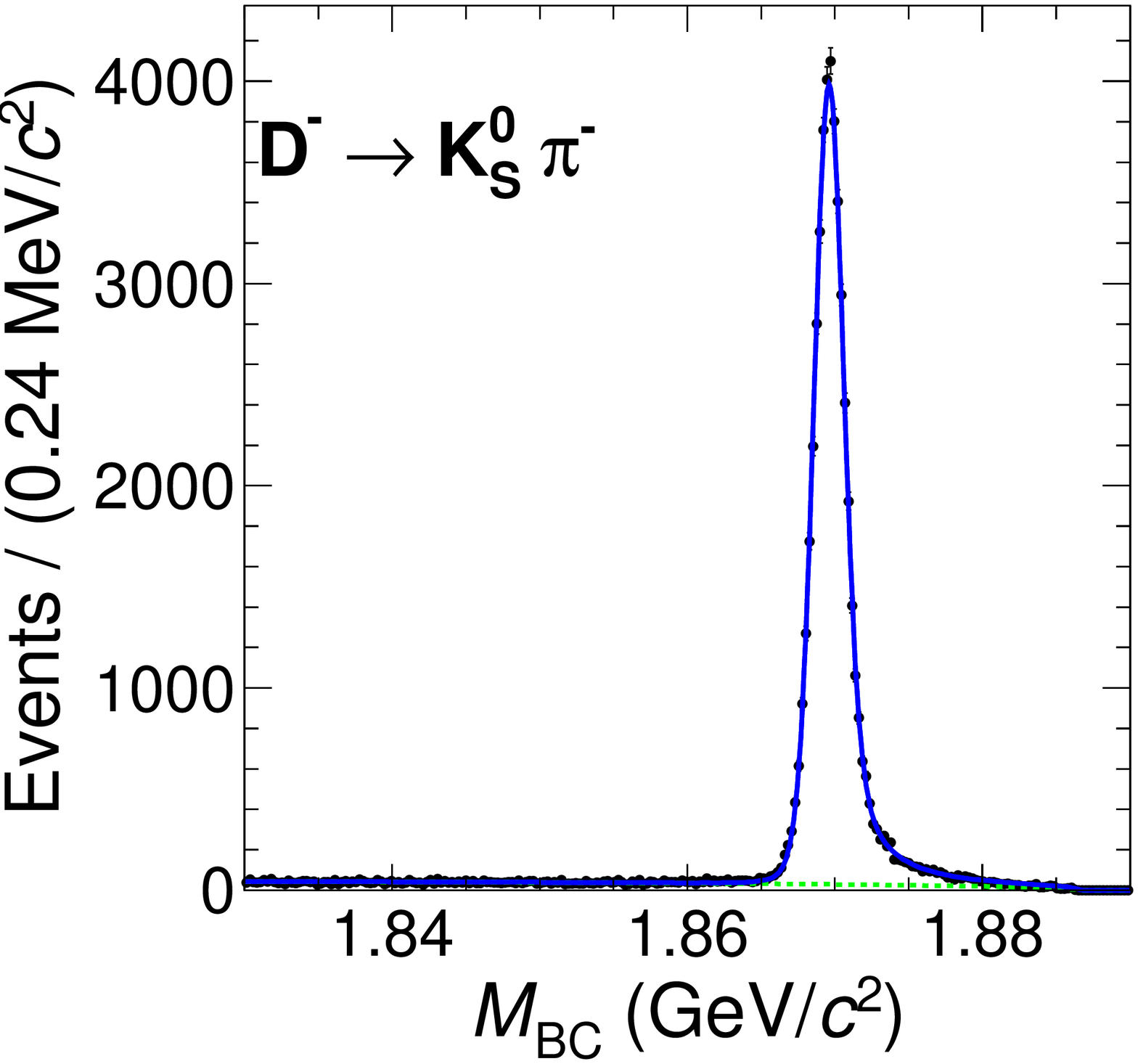}
    \includegraphics[height=3.5cm]{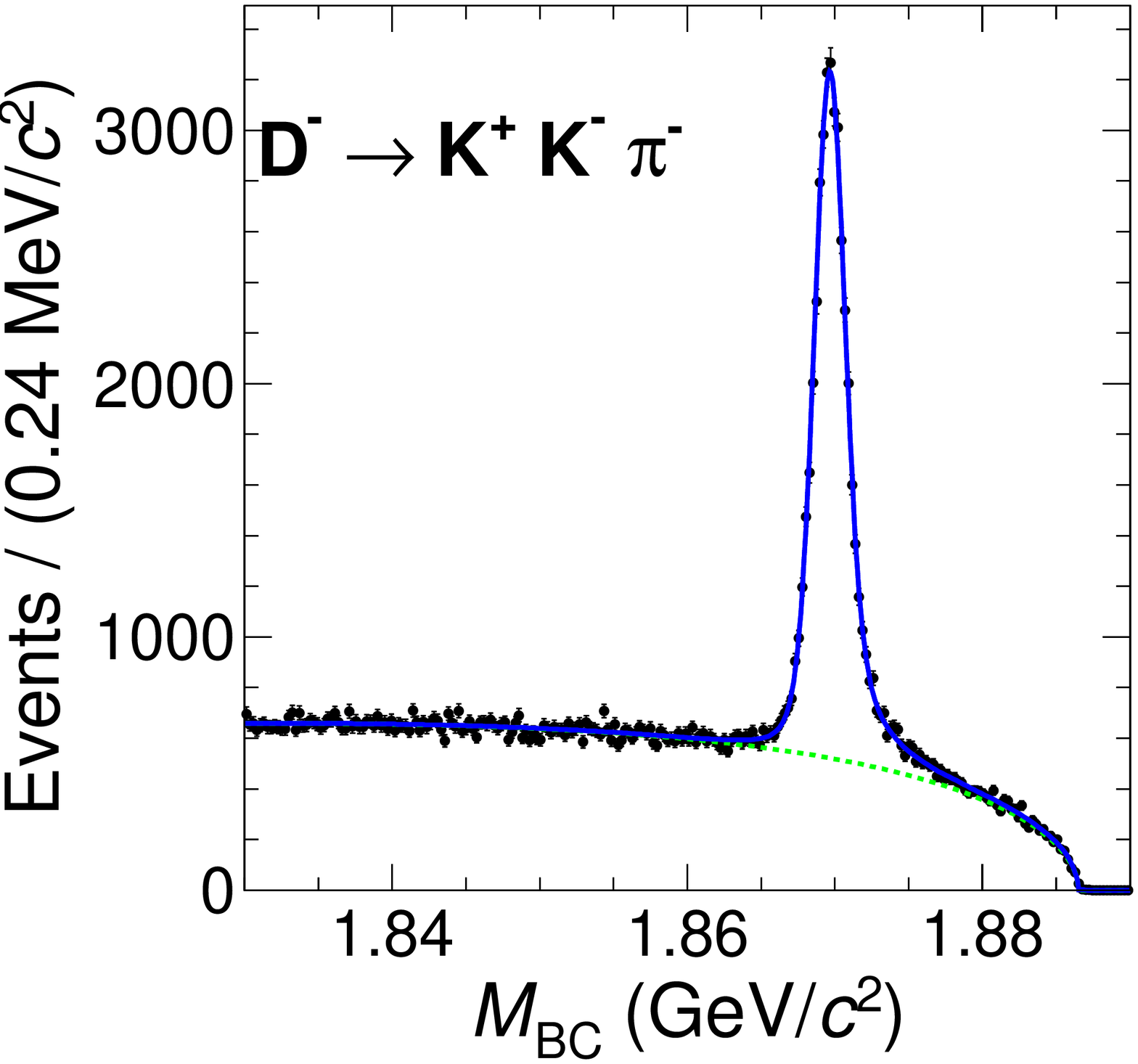}
	\caption{\label{fig:st-mbc-data-dm}Fits to the $\mbc$
	distributions of the ST $D^-$ candidates for data. The dots
	with error bars are data, and the blue solid curves are the
	results of the fits. The green dashed curves are the fitted
	backgrounds.}
\end{figure*}

The charge-conjugated tag modes are fitted simultaneously, with the
same signal and ARGUS background shapes for the tag and charge
conjugated modes.  The numbers of signal and background events are
left free. Figures~\ref{fig:st-mbc-data-dp} and
~\ref{fig:st-mbc-data-dm} show the fits to the $\mbc$ distributions of
the ST $D^+$ and $D^-$ candidates in data, respectively. The ST yields
are obtained by integrating the fitted signal function in the narrower
$\mbc$ signal region ($1.86 < \mbc < 1.88 \GeV/c^2$) and are listed in
Table~\ref{table:brfrac}.

\subsection{Selection of DT events}
\label{subsec:signal}
After ST $D$ candidates are identified, we search for electrons and
$\kl$ showers among the unused charged tracks and neutral
showers. For electron identification, the ratio
$\mathcal{R'}_{\mathcal{L'}}(e)\equiv \mathcal{L'}(e)/[\mathcal{L'}(e)
+ \mathcal{L'}(\pi) + \mathcal{L'}(K)]$ is required to be greater than
0.8, where the likelihood $\mathcal{L'}(i)$ for the hypothesis $i= e$,
$\pi$ or $K$ is formed by combining the EMC information with the
$dE/dx$ and TOF information. The energy lost by electrons to
bremsstrahlung photons is partially recovered by adding the energy of showers
that are within $5^{\circ}$ of the electron and are not
matched to other charged particles. The selected electron is required
to have the opposite charge from the ST $D$. Events that include
charged tracks other than those of the ST $D$ and the electron are
vetoed.

Because of the long $\kl$ lifetime, very few $\kl$ decay in the
MDC. However, most $\kl$ will interact in the material of the EMC,
which gives their position, and deposit part of their energy.
We search for $\kl$ candidates by reconstructing all other particles in
the event;  we then loop over unused reconstructed neutral showers,
taking the direction to the shower as the flight direction of the $\kl$.
Using energy-momentum conservation and the constraint $U_\mathrm{miss}
= 0$, we calculate the momentum magnitude $|\vec{p}_{\kl}|$ of the
$\kl$ and the four-vector of the unreconstructed neutrino in the event.
The variable $U_{\mathrm{miss}}$ is expected to peak at zero for semileptonic
decay candidates and is defined as
\begin{equation}
    U_{\mathrm{miss}} \equiv E_{\mathrm{miss}} - c|\vec{p}_{\mathrm{miss}}|,
\end{equation}
where
\begin{equation}
    \begin{split}
    E_{\mathrm{miss}} &= E_{\mathrm{tot}} - E_{\mathrm{tag}} - E_{\kl} - E_{e},  \\
    \vec{p}_{\mathrm{miss}} &= \vec{p}_{\mathrm{tot}} - \vec{p}_{\mathrm{tag}} - \vec{p}_{\kl} - \vec{p}_{e};
    \end{split}
\end{equation}
$E_{\mathrm{tot}}$, $E_{\mathrm{tag}}$, $E_{\kl}$ and $E_{e}$ are the
energies of the $e^+e^-$, the ST $D$, the $\kl$ and the electron;
$\vec{p}_{\mathrm{tot}}$, $\vec{p}_{\mathrm{tag}}$, $\vec{p}_{\kl}$
and $\vec{p}_{e}$ refer to their momenta. $E_{\kl}$ is calculated by
$E_{\kl} = \sqrt{|\vec{p}_{\kl}|^2 + m_{\kl}^2}$. In order to suppress
background from fake photons, the energy of $\kl$ shower should be
greater than $0.1 \GeV$. We also reject photons that may come from
$\piz$'s by rejecting $\gamma$ in any $\gamma \gamma$ combination
with $0.110 < M_{\gamma \gamma} < 0.155 \GeV/c^2$. In events with
multiple $\kl$ shower candidates, the most energetic shower is
chosen. The inferred four-momentum of the $\kl$ is used to determine
the reconstructed $q^2$, the invariant mass squared of the $e^+ \nu_e$ pair, by
\begin{equation}
    q^2 = \frac{1}{c^4}(E_{\mathrm{tot}} - E_{\mathrm{tag}} - E_{\kl})^2 - \frac{1}{c^2}|\vec{p}_{\mathrm{tot}} - \vec{p}_{\mathrm{tag}} - \vec{p}_{\kl}|^2.
\end{equation}

\begin{figure*}[htbp]
	\centering
    \includegraphics[height=3.5cm]{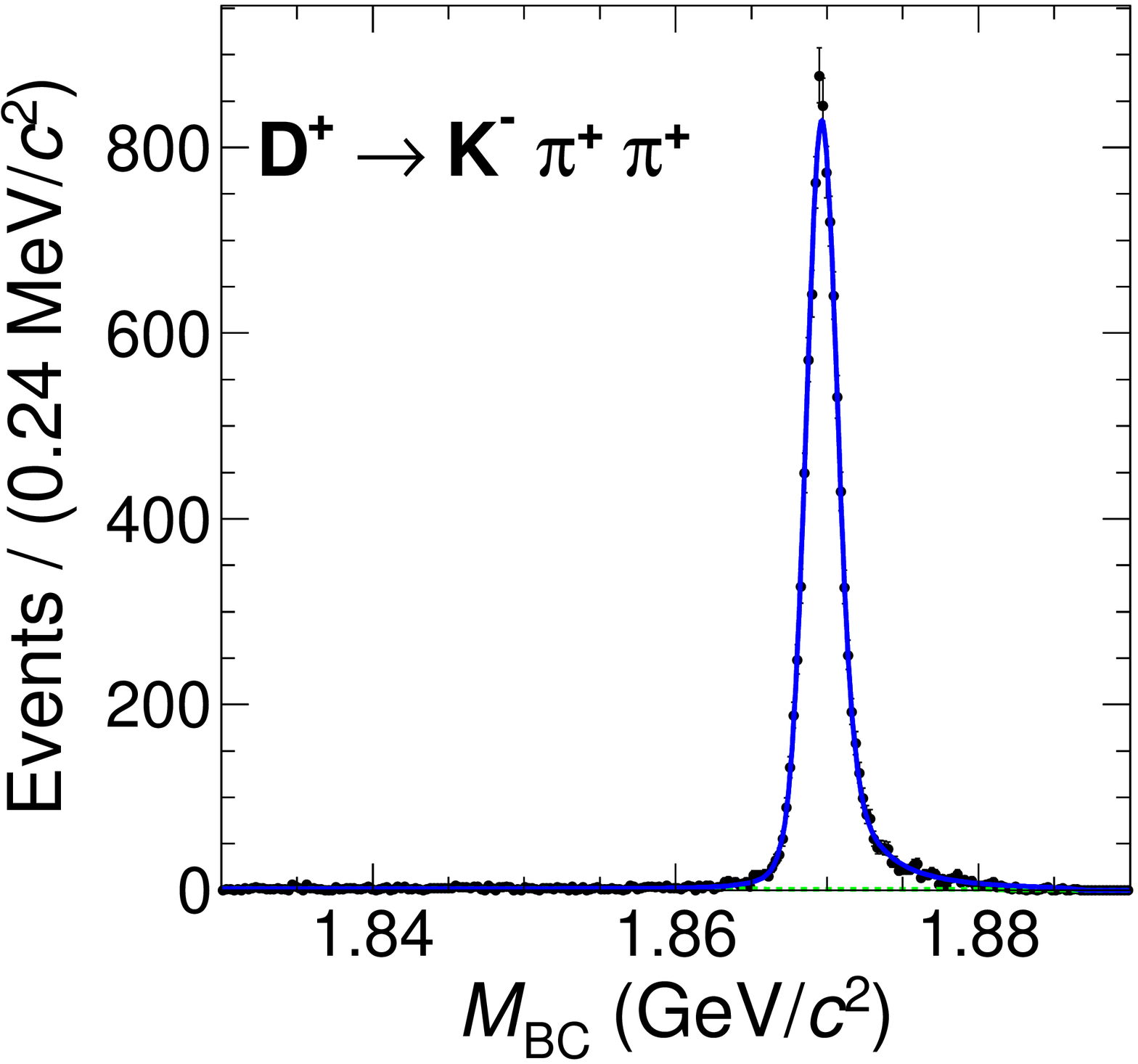}
    \includegraphics[height=3.5cm]{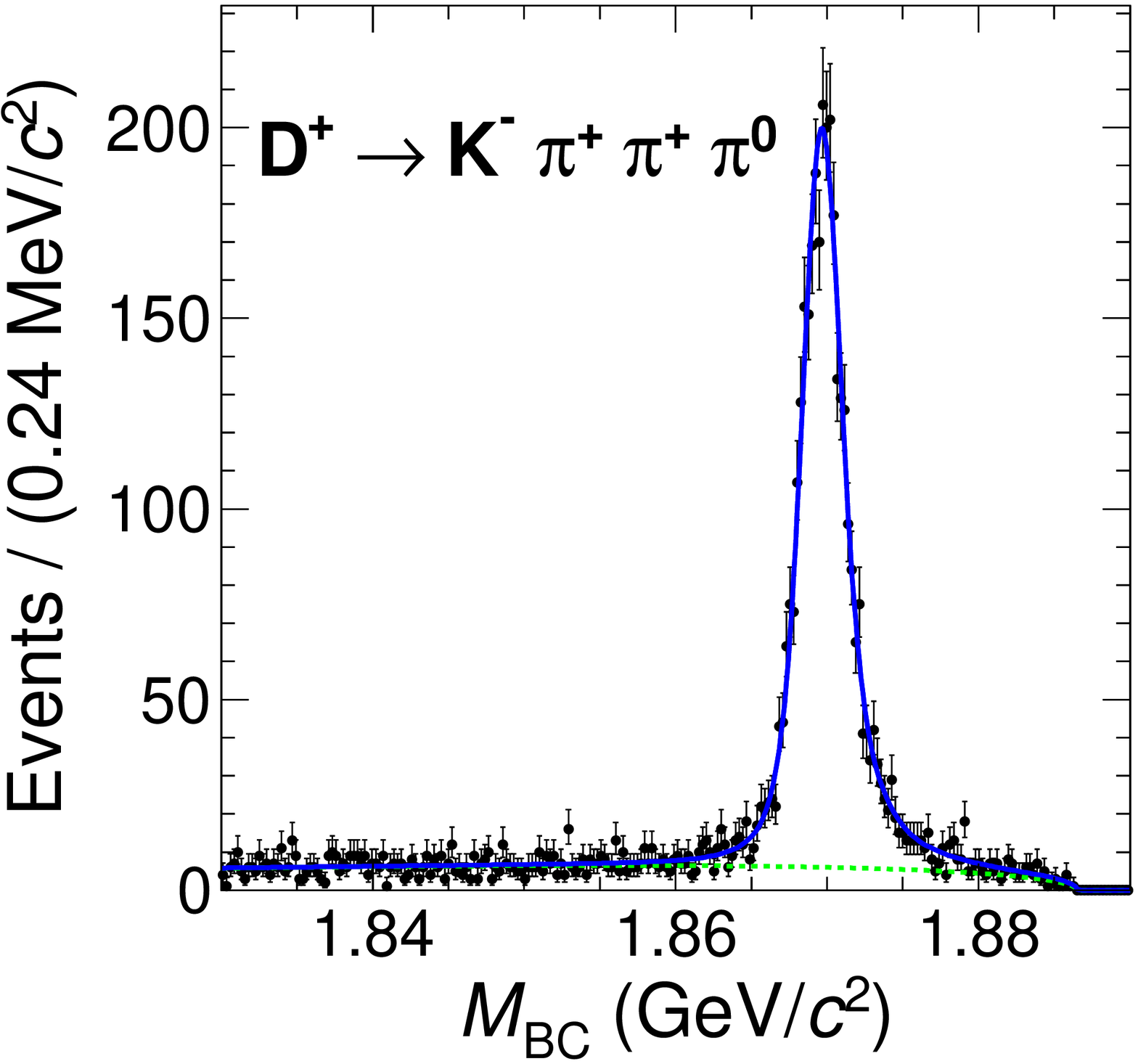}
    \includegraphics[height=3.5cm]{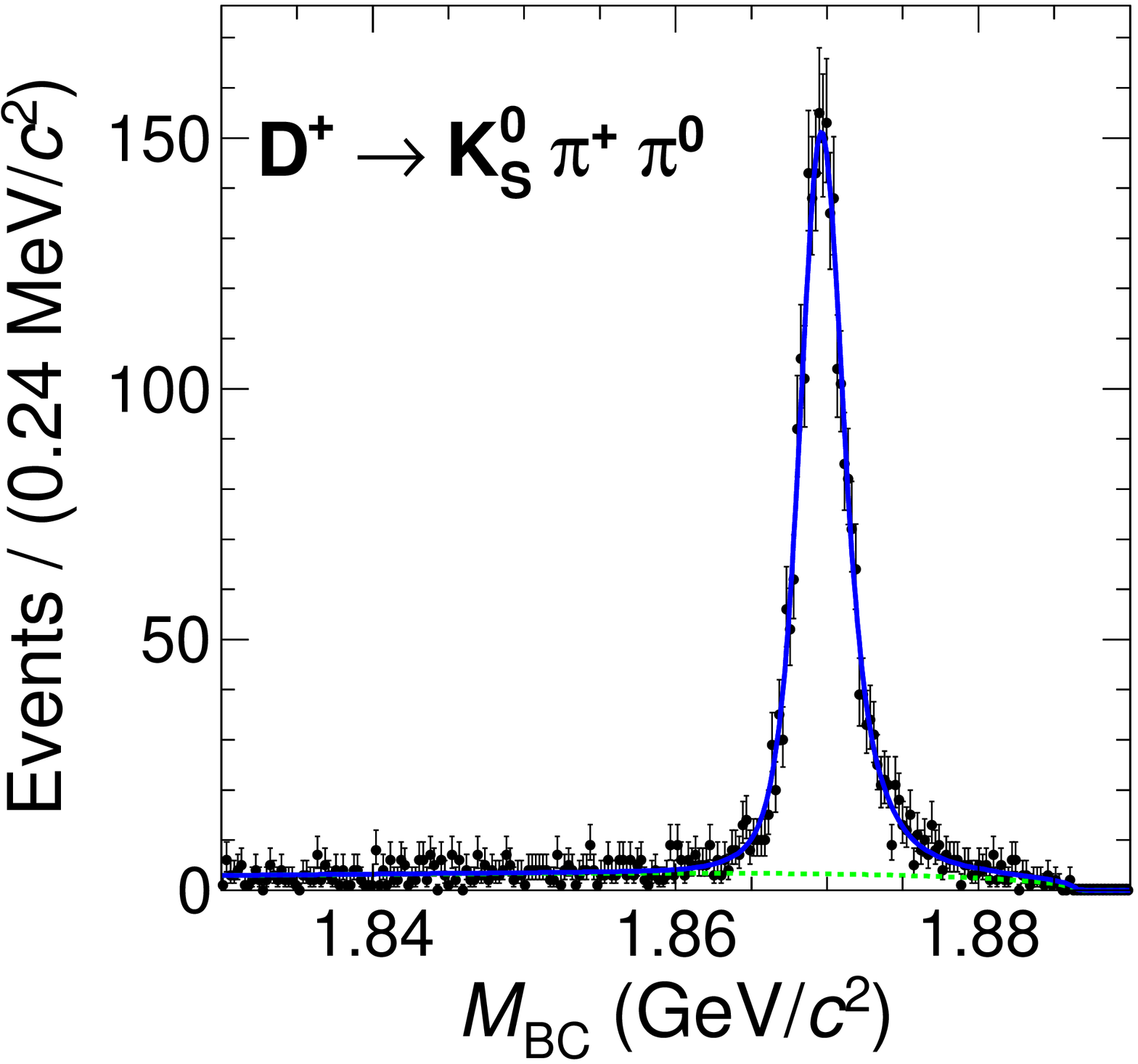}\\
    \includegraphics[height=3.5cm]{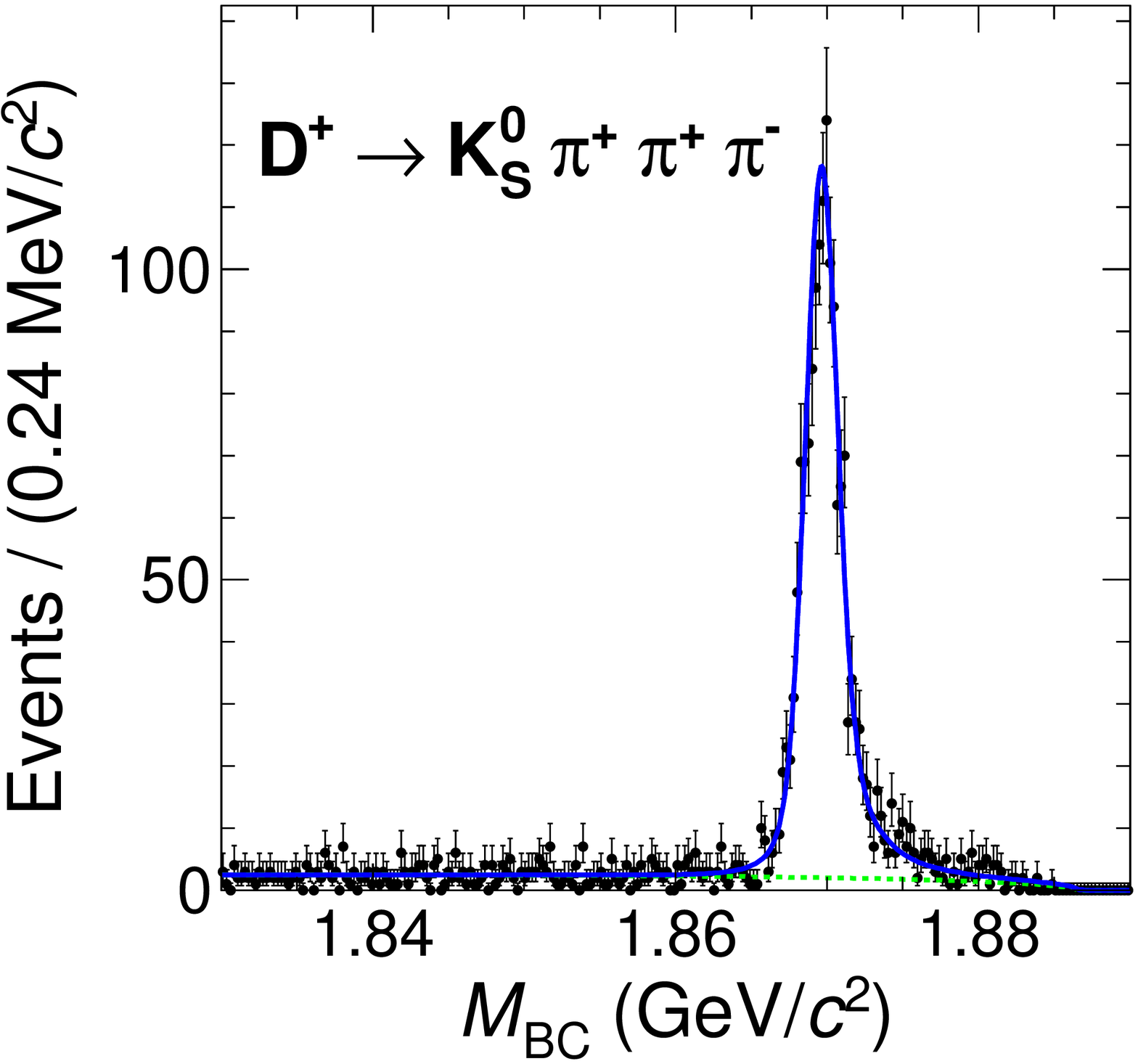}
    \includegraphics[height=3.5cm]{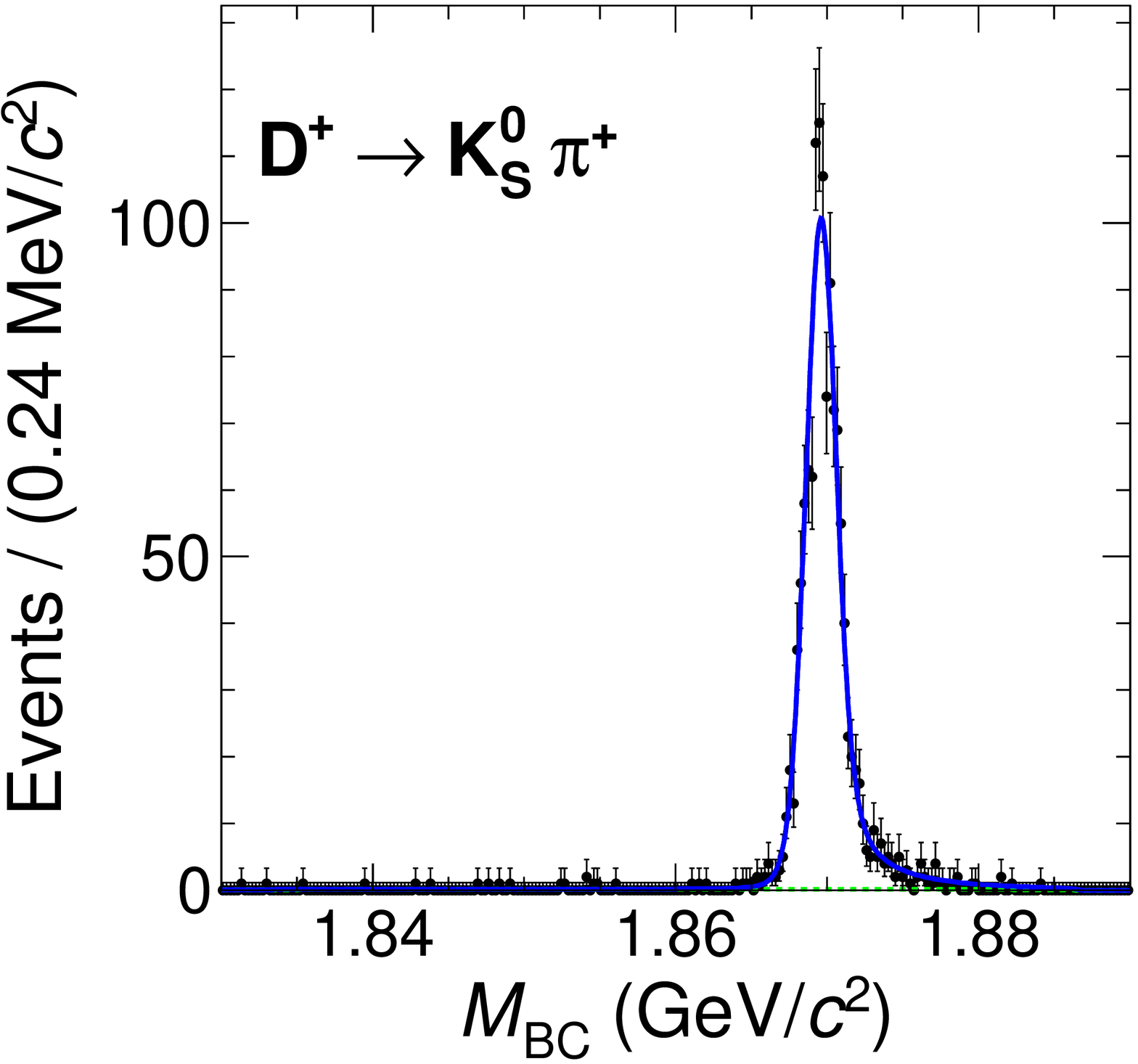}
    \includegraphics[height=3.5cm]{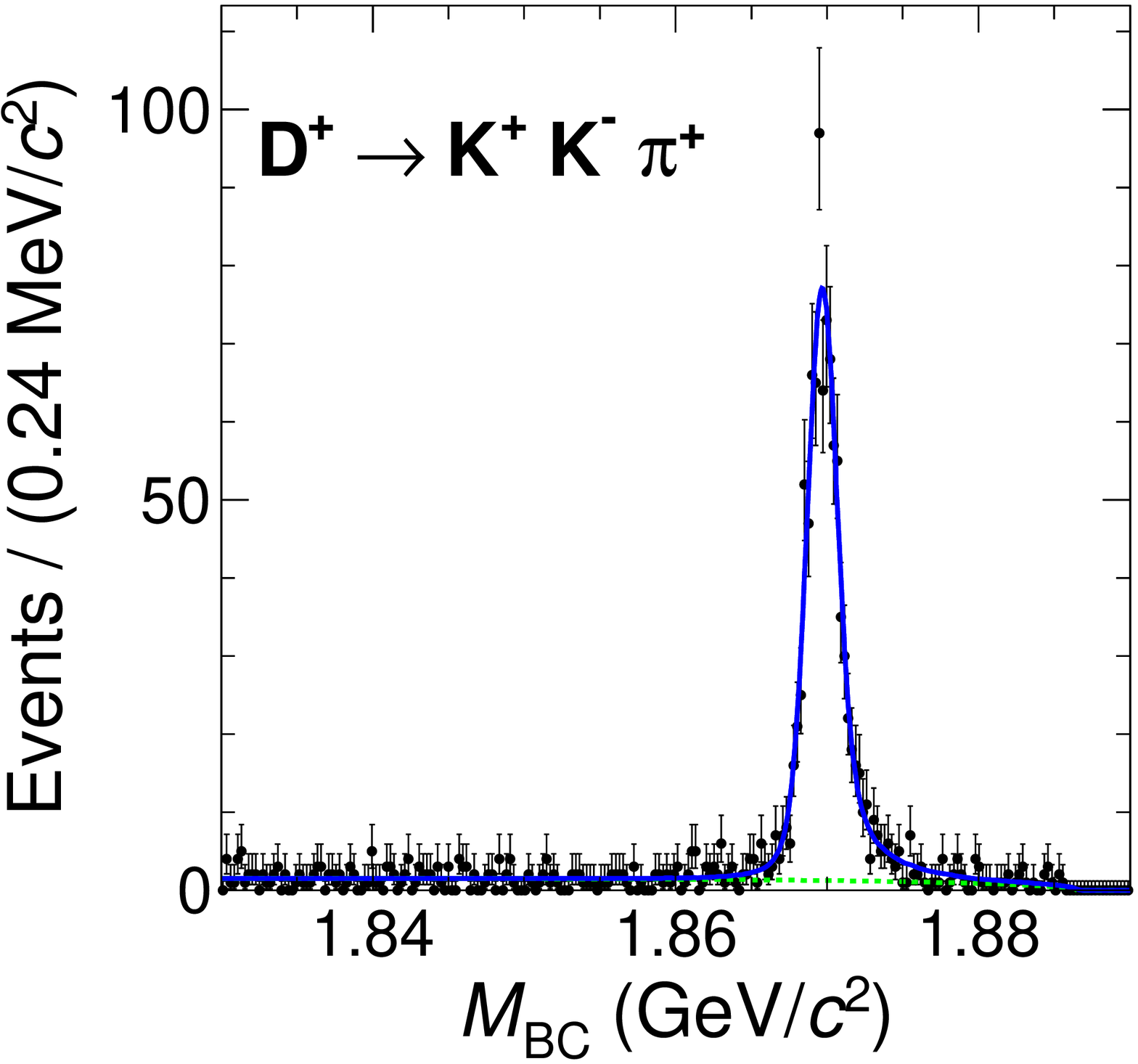}
	\caption{\label{fig:dt-mbc-data-dp}Fits to the $\mbc$
	distributions of the DT $D^+$ candidates for data. The dots
	with error bars are for data, and the blue solid curves are the
	results of the fits. The green dashed curves are the fitted
	combinatorial backgrounds.}
\end{figure*}

\begin{figure*}[htbp]
	\centering
    \includegraphics[height=3.5cm]{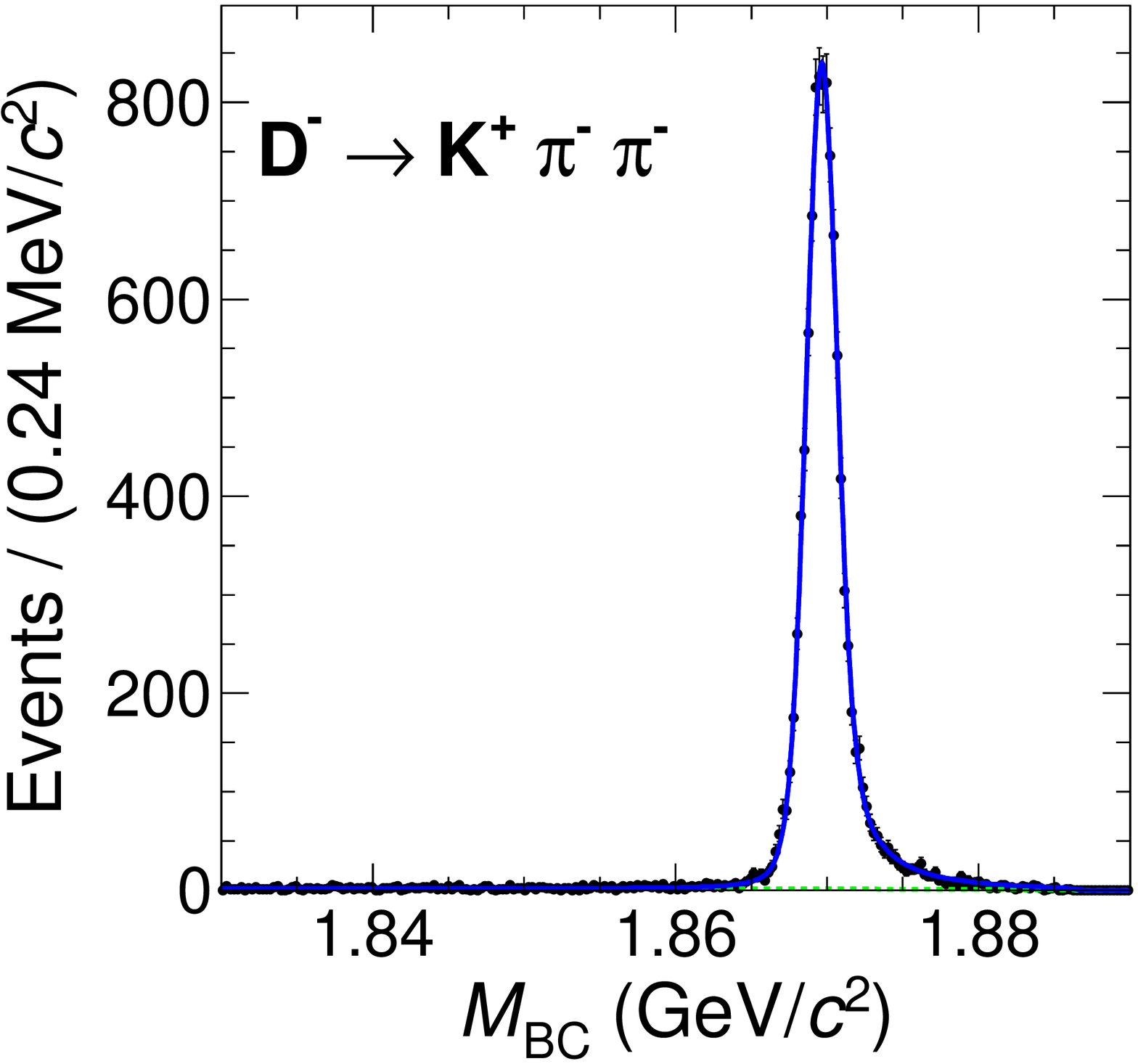}
    \includegraphics[height=3.5cm]{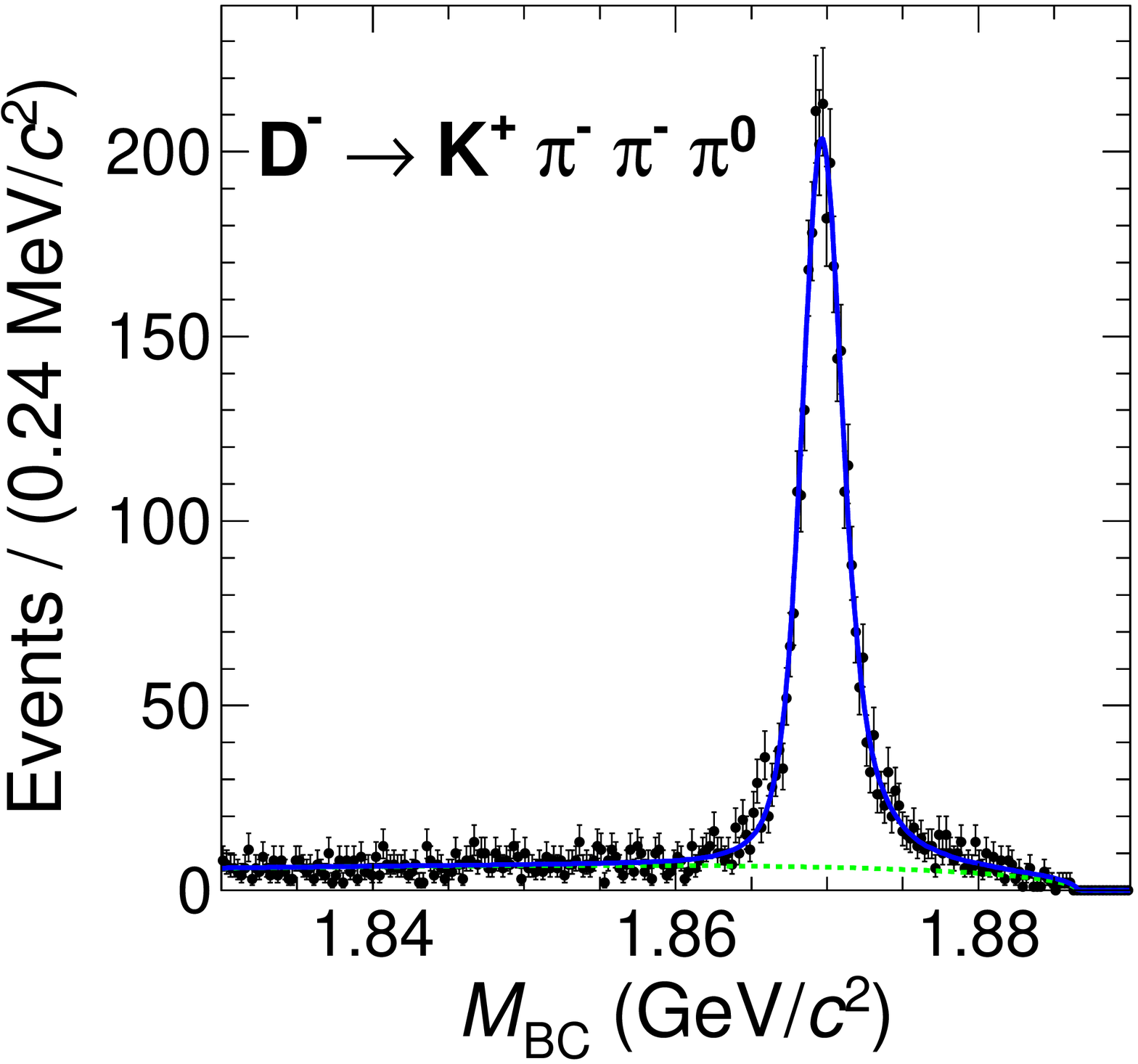}
    \includegraphics[height=3.5cm]{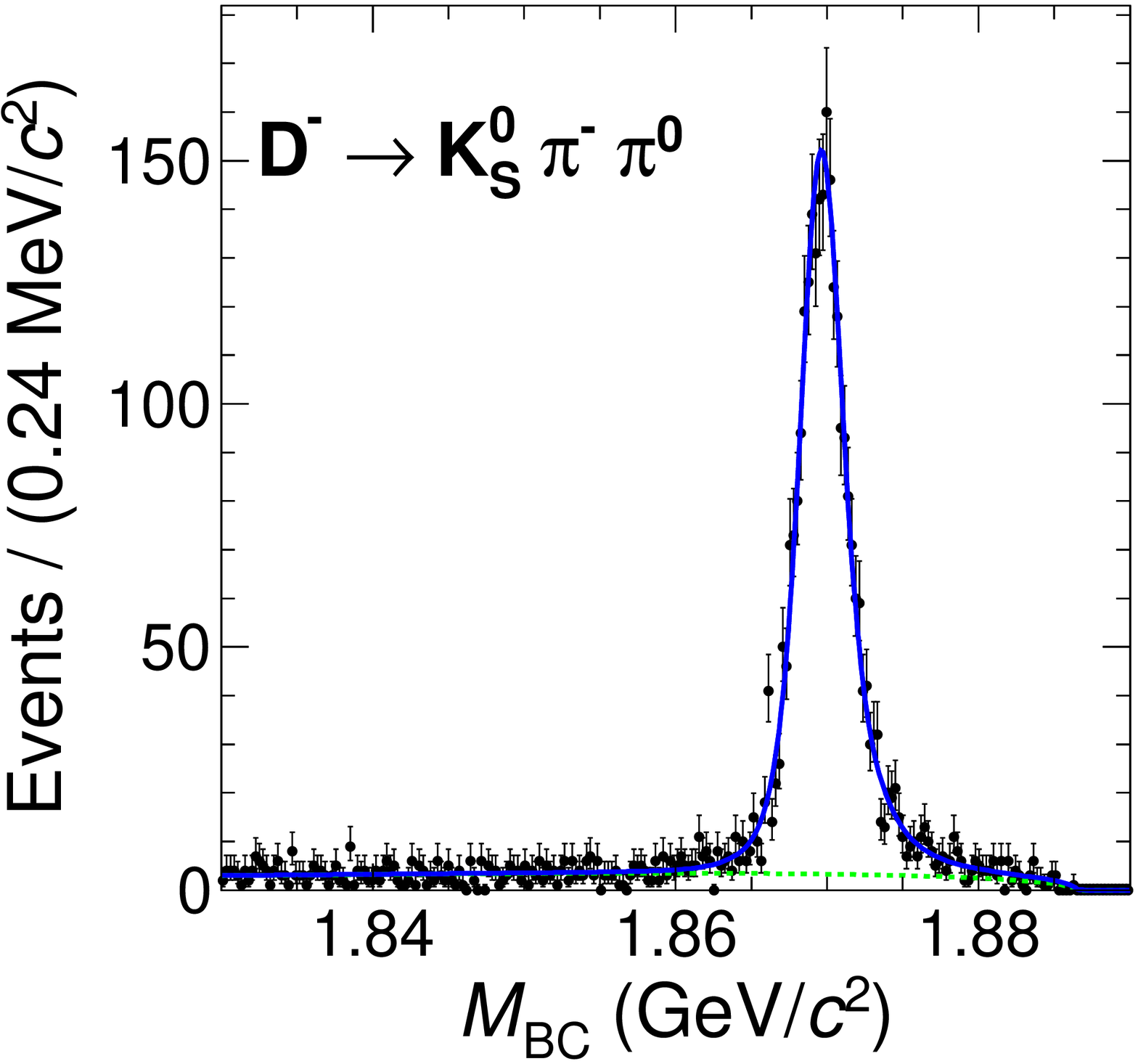}\\
    \includegraphics[height=3.5cm]{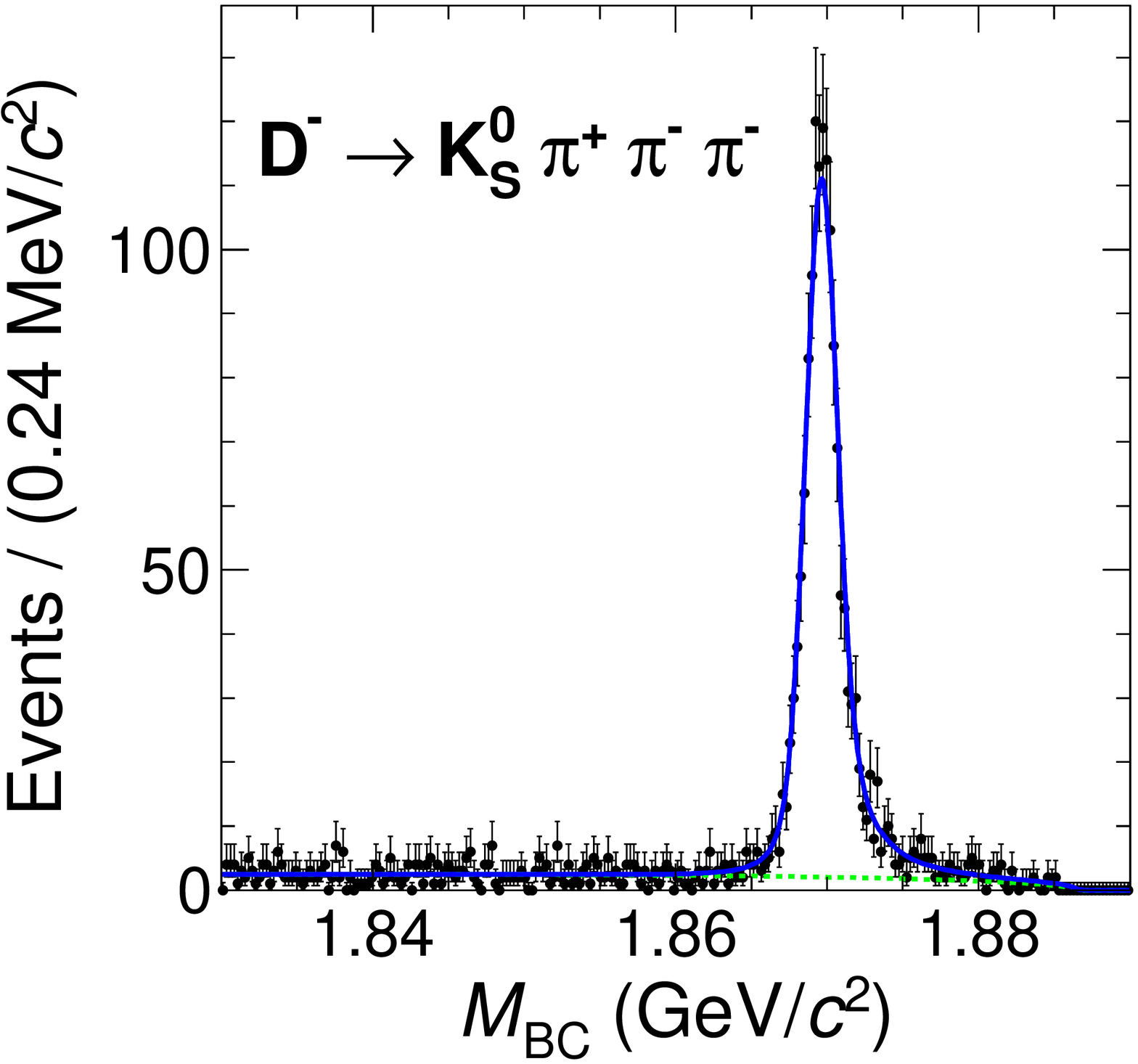}
    \includegraphics[height=3.5cm]{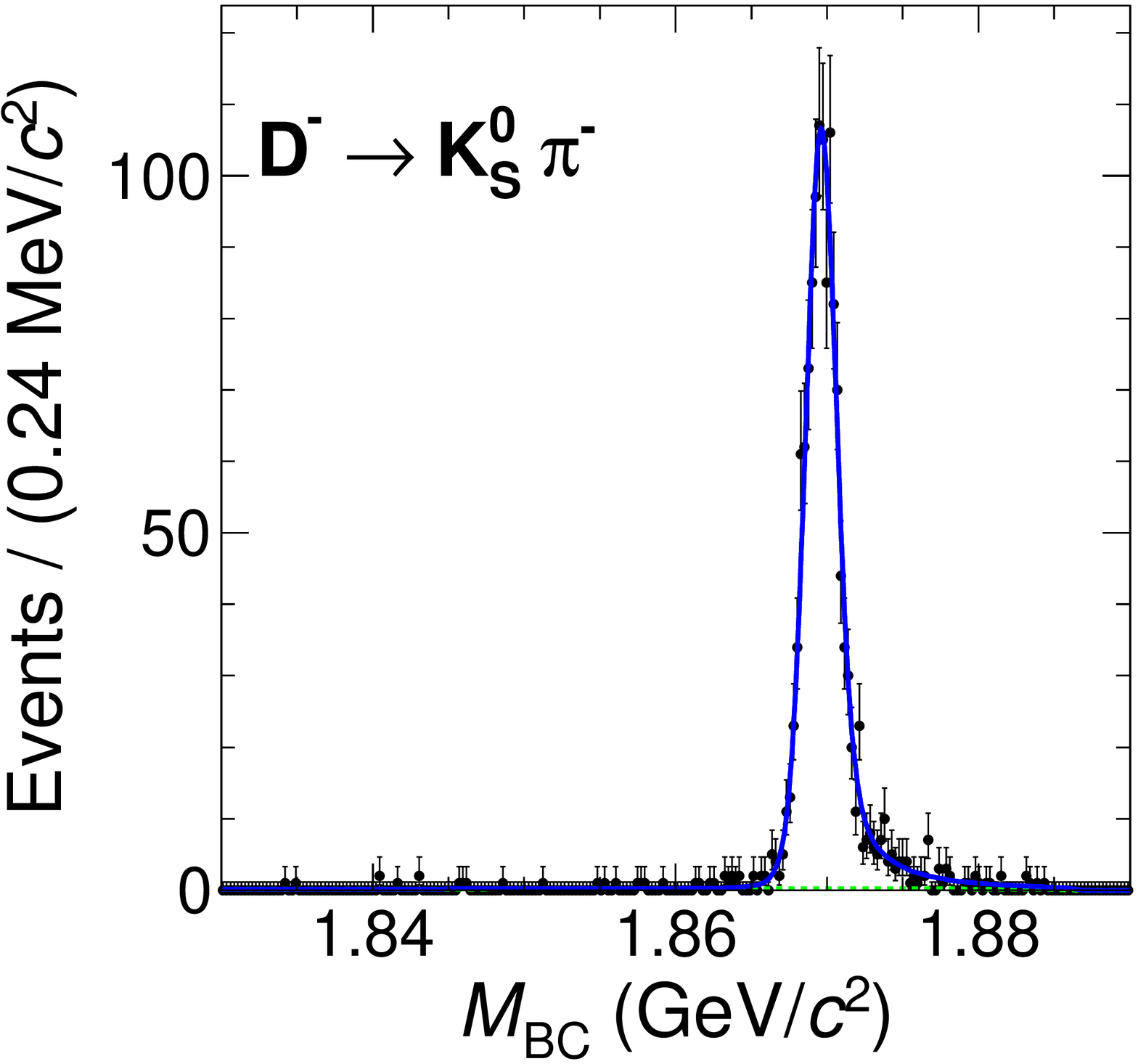}
    \includegraphics[height=3.5cm]{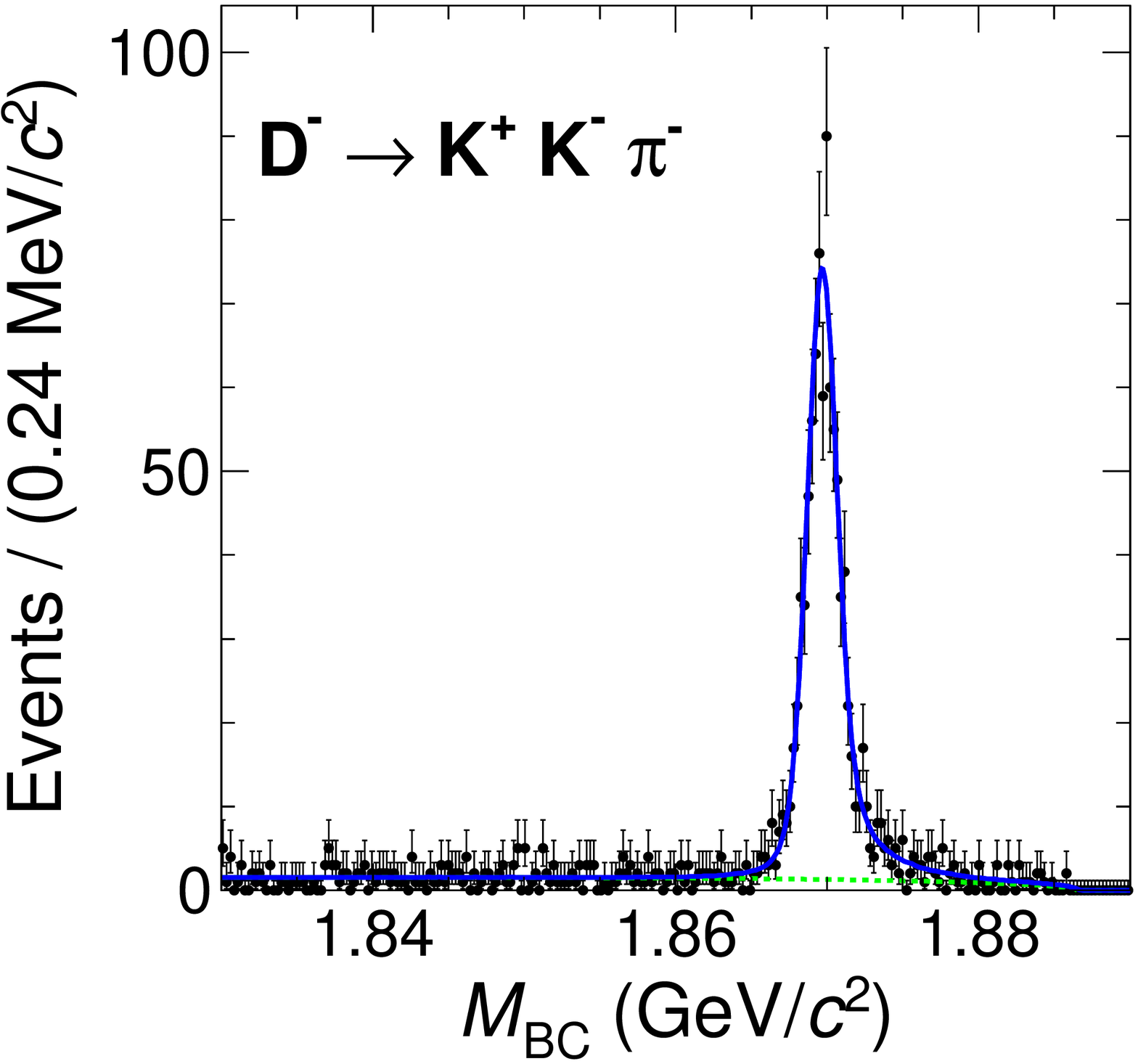}
	\caption{\label{fig:dt-mbc-data-dm}Fits to the $\mbc$
	distributions of the DT $D^-$ candidates for data. The dots
	with error bars are for data, and the blue solid curves are the
	results of the fits. The green dashed curves are the fitted
	combinatorial backgrounds.}
\end{figure*}

Similar to the determination of the ST yields, we obtain the DT yields
of data from the fits to the $\mbc$ distributions of the corresponding
ST $D$ candidates. Figures~\ref{fig:dt-mbc-data-dp} and
~\ref{fig:dt-mbc-data-dm} show the fits to the $\mbc$ distributions of
the DT $D^+$ and $D^-$ candidates in data, respectively. From the
fits, we obtain the DT yields in data, which are listed in the third
column of Table~\ref{table:brfrac}.

\subsection{Estimation of backgrounds}
\label{subsec:peakbkg}
The $\kl$ reconstruction efficiencies of data and MC differ, so the
$\kl$ reconstruction efficiency of the generic MC is corrected to that
of data. The correction factors of $\kl$ reconstruction efficiencies
are determined from two control samples ($\jpsi \to K^{*}(892)^{\pm}
K^{\mp}$ with $K^*(892)^{\pm} \to \kl \pi^{\pm}$ and $\jpsi \to \phi
\kl K^{\pm} \pi^{\mp}$), which are described in
Appendix~\ref{app:kl-sys}. The corrected generic MC samples are used
to determine the amount of peaking background and the efficiency for
$D^+ \to \kl e^+ \nu_e$.

We examine the topologies of the corrected generic MC samples to study
the composition of the DT samples. In the $\mbc$ signal region, the
DT $D$ candidates can be divided into the following categories:
\bitm
    \item Signal: Tag-side and signal-side correctly matched.
    \item Background:
    \bitm
        \item Tag-side mismatched events (Bkg~\RN{1}).
        \item Tag-side matched but signal-side mismatched signal events (Bkg~\RN{2}).
        \item Tag-side matched but $D \to X e \nu_e$ non-signal events on signal side (Bkg~\RN{3}).
        \item Tag-side matched but $D \to X \mu \nu_{\mu}$ events on signal side (Bkg~\RN{4}).
        \item Tag-side matched but non-leptonic $D$ decay events on signal side (Bkg~\RN{5}).
    \eitm
\eitm

In the selected DT candidates, the proportion of signal events varies
from $49\%$ to $58\%$ according to the specific hadronic tag
mode. Bkg~\RN{1} comes from $\ddb$ decays in which the hadronic tag
$D$ is mis-reconstructed and non-$\ddb$ processes, and varies from
$1\%$ to $12\%$ according to the specific hadronic tag
mode. Bkg~\RN{2} ($\sim$10$\%$) consists of $D^+ \to \kl e^+ \nu_e$
events of which $\kl$ shower is mis-reconstructed. The dominant
background in the DT sample is Bkg~\RN{3} ($\sim$24$\%$), which is
from $D^+ \to \bar K^{*}(892)^{0} e^+ \nu_e$ ($41.9\%$), $D^+ \to \ks
e^+ \nu_e$ ($41.2\%$), $D^+ \to \pi^0 e^+ \nu_e$ ($10.2\%$), $D^+ \to
\eta e^+ \nu_e$ ($6.0\%$) and $D^+ \to \omega e^+ \nu_e$
($0.7\%$). Bkg~\RN{4} ($\sim$3$\%$) consists of $D^+ \to \kl \mu^+
\nu_\mu$ ($65.2\%$), $D^+ \to \bar K^{*}(892)^{0} \mu^+ \nu_\mu$
($23.3\%$) and $D^+ \to \ks \mu^+ \nu_\mu$ ($11.5\%$). Bkg~\RN{5}
($\sim$3$\%$) consists of $D^+ \to \bar K^0 \pi^+ \pi^0$ ($78\%$) and
$D^+ \to \bar K^0 K^{*}(892)^+$ ($22\%$).

\section{Branching Fraction and $CP$ Asymmetry}
\label{sec:brfraction}

The branching fraction for $D^+ \to \kl e^+ \nu_e$
($\mathcal{B}_{\mathrm{sig}}$) is determined by
\begin{equation}
    \mathcal{B}_{\mathrm{sig}} = \frac{N_{\mathrm{DT}}(1 - f_{\mathrm{bkg}}^{\mathrm{peak}})}{\eff N_{\mathrm{ST}}},
     \label{eqn:br-calculation}
\end{equation}
where $N_{\mathrm{DT}}$, $N_{\mathrm{ST}}$ are the DT and
ST yields, $f_{\mathrm{bkg}}^{\mathrm{peak}}$ is the
proportion of peaking backgrounds in the DT candidates (from
Bkg~\RN{2} to Bkg~\RN{5}), $\eff$ is the efficiency for
finding $D^+ \to \kl e^+ \nu_e$ in the presence of ST
$D$. $f_{\mathrm{bkg}}^{\mathrm{peak}}$ and $\eff$ are obtained from
the $\kl$ efficiency corrected generic MC samples. The
$D^+ \to \kl e^+ \nu_e$  branching fractions for
different ST modes are listed in Table~\ref{table:brfrac}. We obtain
$\mathcal{B}(D^+ \to \kl e^+ \nu_e) = (4.454 \pm 0.038 \pm 0.102)\%$
and $\mathcal{B}(D^- \to \kl e^- \bar \nu_e) = (4.507 \pm 0.038 \pm
0.104)\%$, which are the weighted averages of the six ST modes for
$D^+$ and $D^-$ separately. Combining these branching fractions, we
obtain the averaged branching fraction $\mathcal{\bar B}(D^+ \to \kl
e^+ \nu_e) = (4.481 \pm 0.027 \pm 0.103)\%$, which agrees well with
the measurement of $\mathcal{B}(D^+ \to \ks e^+ \nu_e)$ of
CLEO-$c$~\cite{besson}. The $CP$ asymmetry of $D^+ \to \kl e^+ \nu_e$ is
\begin{equation}
    \begin{split}
    A_{CP} &\equiv \frac{\mathcal{B}(D^+ \to \kl e^+ \nu_e) - \mathcal{B}(D^- \to \kl e^- \bar \nu_e)}{\mathcal{B}(D^+ \to \kl e^+ \nu_e) + \mathcal{B}(D^- \to \kl e^- \bar \nu_e)} \\
    &= (-0.59 \pm 0.60 \pm 1.48)\%.
    \end{split}
\end{equation}
This result is consistent with the theoretical prediction in
Ref.~\cite{xingzz} ($-3.3 \times 10^{-3}$).

\begin{table*}[htbp]
	\centering
        \caption{\label{table:brfrac} Summary of the ST yields ($N_{\mathrm{ST}}$), the DT yields ($N_{\mathrm{DT}}$), the peaking background rates for the DT candidates ($f_{\mathrm{bkg}}^{\mathrm{peak}}$), the detection efficiency ($\eff$) and the branching fraction for signal decay for each ST mode ($\mathcal{B}_{\mathrm{sig}}$). The averages are the weighted average of the individual ST mode branching fractions. The uncertainties are statistical.}
	\begin{tabular}{l|ccccc}\hline\hline
        \multicolumn{6}{c}{$D^+ \to \kl e^+ \nu_e$}   \\
        \hline
		\multicolumn{1}{c|}{Tag Mode}                 &
		\multicolumn{1}{c}{$N_{\mathrm{ST}}$}         &
		\multicolumn{1}{c}{$N_{\mathrm{DT}}$}         &
		\multicolumn{1}{c}{$f_{\mathrm{bkg}}^{\mathrm{peak}} (\%)$} &
		\multicolumn{1}{c}{$\eff (\%)$}               &
		\multicolumn{1}{c}{$\mathcal{B}_{\mathrm{sig}} (\%)$} \\
		\hline
		$D^{-} \rightarrow K^+ \pi^- \pi^-$       & $410200 \pm  670$ & $10492\pm  103$ & $41.83 \pm 0.28$ &  $33.96 \pm 0.10$ & $4.381 \pm 0.050$\\
		$D^{-} \rightarrow K^+ \pi^- \pi^- \piz$  & $120060 \pm  457$ & $3324 \pm   64$ & $44.78 \pm 0.49$ &  $33.14 \pm 0.19$ & $4.613 \pm 0.103$\\
		$D^{-} \rightarrow \ks \pi^- \piz$        & $102136 \pm  378$ & $2658 \pm   56$ & $38.93 \pm 0.58$ &  $35.67 \pm 0.21$ & $4.456 \pm 0.108$\\
		$D^{-} \rightarrow \ks \pi^- \pi^- \pi^+$ & $ 59158 \pm  303$ & $1459 \pm   41$ & $40.84 \pm 0.76$ &  $32.51 \pm 0.27$ & $4.488 \pm 0.145$\\
		$D^{-} \rightarrow \ks \pi^-$             & $ 47921 \pm  225$ & $1287 \pm   36$ & $38.90 \pm 0.88$ &  $35.07 \pm 0.32$ & $4.679 \pm 0.155$\\
		$D^{-} \rightarrow K^+ K^- \pi^-$         & $ 35349 \pm  239$ & $ 905 \pm   32$ & $44.64 \pm 0.97$ &  $30.98 \pm 0.35$ & $4.575 \pm 0.190$\\
		\hline
		Average & & & & & $4.454 \pm 0.038$\\
		\hline
        \multicolumn{6}{c}{$D^- \to \kl e^- \bar \nu_e$}   \\
        \hline
		\multicolumn{1}{c|}{Tag Mode}                 &
        \multicolumn{1}{c}{$N_{\mathrm{ST}}$}         &
		\multicolumn{1}{c}{$N_{\mathrm{DT}}$}         &
		\multicolumn{1}{c}{$f_{\mathrm{bkg}}^{\mathrm{peak}} (\%)$} &
		\multicolumn{1}{c}{$\eff (\%)$}               &
		\multicolumn{1}{c}{$\mathcal{B}_{\mathrm{sig}} (\%)$} \\
		\hline
		$D^{+} \rightarrow K^- \pi^+ \pi^+$       & $407666 \pm  668$ & $10354\pm  103$ & $40.44 \pm 0.29$ &  $34.02 \pm 0.11$ & $4.447 \pm 0.051$\\
		$D^{+} \rightarrow K^- \pi^+ \pi^+ \piz$  & $117555 \pm  450$ & $3264 \pm   63$ & $42.28 \pm 0.52$ &  $33.19 \pm 0.19$ & $4.829 \pm 0.107$\\
		$D^{+} \rightarrow \ks \pi^+ \piz$        & $101824 \pm  378$ & $2642 \pm   55$ & $39.06 \pm 0.58$ &  $35.92 \pm 0.21$ & $4.402 \pm 0.104$\\
		$D^{+} \rightarrow \ks \pi^+ \pi^+ \pi^-$ & $ 59046 \pm  303$ & $1533 \pm   42$ & $39.68 \pm 0.77$ &  $33.44 \pm 0.27$ & $4.683 \pm 0.147$\\
		$D^{+} \rightarrow \ks \pi^+$             & $ 48240 \pm  226$ & $1217 \pm   35$ & $38.50 \pm 0.88$ &  $35.20 \pm 0.32$ & $4.408 \pm 0.147$\\
		$D^{+} \rightarrow K^+ K^- \pi^+$         & $ 35742 \pm  240$ & $ 942 \pm   32$ & $44.04 \pm 0.95$ &  $32.40 \pm 0.36$ & $4.552 \pm 0.181$\\
		\hline
		Average & & & & & $4.507 \pm 0.038$\\
		\hline\hline
	\end{tabular}
\end{table*}

Table~\ref{table:sys-br} summarizes the systematic uncertainties in
the measurements of absolute branching fractions and the $CP$
asymmetry of $D^+ \to \kl e^+ \nu_e$. A brief description of each
systematic uncertainty is provided below.

\begin{table*}[htbp]
	\centering
        \caption{\label{table:sys-br}Systematic uncertainties in the measurements of absolute branching fraction and the $CP$ asymmetry of $D^+ \to \kl e^+ \nu_e$.}
	\begin{tabular}{lcc}
		\hline\hline
		Source & $D^+ \to \kl e^+ \nu_e (\%)$ & $D^- \to \kl e^- \bar \nu_e (\%)$ \\
        \hline
        Electron tracking                                     & $0.5$   & $0.5$ \\
        Electron ID                                           & $0.1$   & $0.1$ \\
        $\kl$ efficiency correction                           & $1.2$   & $1.2$ \\
        Extra $\chi^2$ cut for $\kl$ efficiency correction    & $0.8$   & $0.8$ \\
        Peaking backgrounds in DT                             & $1.6$   & $1.6$ \\
        $\mbc$ fit                                            & negligible & negligible \\
        \hline
        Total (Branching fraction)                            & $2.3$   & $2.3$ \\
        Total ($CP$ asymmetry)                                & $2.1$   & $2.1$ \\
		\hline\hline
	\end{tabular}
\end{table*}

\begin{enumerate}
    \item \emph{Electron (positron) track-finding and identification (ID) efficiency}

      Uncertainties of electron (positron) track-finding and ID
      efficiency are obtained by comparing the track-finding and ID
      efficiencies for the electrons (positrons) from radiative Bhabha
      processes in the data and MC. Considering both the $\cos\theta$,
      where $\theta$ is the polar angle of the positron, and momentum
      distributions of the electrons (positrons) of the signal events,
      we obtain the two-dimensional weighted uncertainty of electron
      (positron) track-finding to be $0.5\%$, and the averaged
      uncertainties of positron and electron ID efficiency to be
      $0.03\%$ and $0.10\%$, respectively.

    \item \emph{$\kl$ efficiency correction}

      We take the relative statistical uncertainty of the $\kl$
      efficiency difference between data and MC as a function of
      momentum (as shown in Fig.~\ref{fig:kl-eff-comparison} in
      Appendix~\ref{app:kl-sys}) as the uncertainty of the $\kl$
      efficiency correction. Weighting these uncertainties by the
      $\kl$ momentum distribution in $D^+ \to \kl e^+ \nu_e$, we
      obtain the uncertainties of the $K^0 \to \kl$ and $\bar K^0 \to
      \kl$ efficiency corrections to both be $1.2\%$.

    \item \emph{Extra $\chi^2$ cut for $\kl$ efficiency correction}

      As described in Appendix~\ref{app:kl-sys}, in the determination
      of correction factor of the $\kl$ efficiency, we apply a
      $\chi^2$ cut which brings an extra uncertainty. The uncertainty
      of the $\chi^2$ cut is obtained by comparing the cut efficiency
      between data and MC using two control samples ($\jpsi \to
      K^{*}(892)^{\pm} K^{\mp}$ with $K^*(892)^{\pm} \to \kl
      \pi^{\pm}$ and $\jpsi \to \phi \kl K^{\pm}
      \pi^{\mp}$). Weighting by the momentum distribution of the $\kl$
      of signal events, the uncertainty of the extra $\chi^2$ cut
      ($\chi^2 < 100$) is $0.8 \%$.

    \item \emph{Peaking backgrounds in DT}

      For Bkg~\RN{2}, from Eq.~(\ref{eqn:br-calculation}) the ratio of
      mis-reconstructed $\kl$ will not affect the measured branching
      fraction, since the numerator and the denominator share the
      common factor. The uncertainties of the peaking backgrounds of
      mis-reconstructed $\kl$ can be safely ignored. For Bkg~\RN{3},
      Bkg~\RN{4} and Bkg~\RN{5}, we determine the change of the number
      of DT events by varying the branching fractions of peaking
      background channels by $1\sigma$, and the uncertainty of peaking
      backgrounds in DT events is $1.6\%$.

    \item \emph{$\mbc$ fit}

      To evaluate the systematic uncertainty from the $\mbc$ fit, we
      determine the changes of the DT yields divided by the ST yields
      when varying the standard deviation of the convoluted Gaussian
      function by $\pm1\sigma$ deviation for each tag mode. We find
      that they are negligible.
\end{enumerate}

The total systematic uncertainties of the branching fractions for $D^+
\to \kl e^+ \nu_e$ and $D^- \to \kl e^- \bar \nu_e$ are determined to
be $2.3\%$ and $2.3\%$, respectively, by adding all contributions in
quadrature. In the determination of the $CP$ asymmetry, the
corresponding systematic uncertainties of branching fractions for $D^+
\to \kl e^+ \nu_e$ and $D^- \to \kl e^- \bar \nu_e$ are obtained in a
similar fashion, except that the contribution of the extra $\chi^2$
cut of $\kl$ efficiency correction is not used since it cancels.
The systematic uncertainties entering the $CP$ asymmetry are found to be $2.1\%$ and $2.1\%$,
respectively.

\section{Hadronic Form Factor}
\label{sec:formfactor}

\subsection{Method of extraction of form factor}
The number of produced signal events for each tag mode from the whole
$q^2$ range can be written as
\begin{equation}
    n = 2N_{D^+ D^-}\mathcal{B}_{\mathrm{tag}}\mathcal{B}_{\mathrm{sig}} = N_{\mathrm{tag}}\frac{\Gamma_{\mathrm{sig}}}{\Gamma_{D^+}},
\end{equation}
where $\Gamma_{\mathrm{sig}}$ is the partial decay width of $D^+ \to
\kl e^+ \nu_e$ while $\Gamma_{D^+}$ is the total decay width of
$D^+$. So we obtain
\begin{equation}
    \label{eqn:dn}
    dn = \frac{N_{\mathrm{tag}}}{\Gamma_{D^+}}d\Gamma_{\mathrm{sig}} = N_{\mathrm{tag}}\tau_{D^+}d\Gamma_{\mathrm{sig}},
\end{equation}
where $\tau_{D^+} = 1/\Gamma_{D^+}$ is the $D^+$ lifetime and $d\Gamma_{\mathrm{sig}}$ is the differential decay width of the signal.

Substituting Eq.~(\ref{eqn:dn}) into Eq.~(\ref{eqn:dgdq2}),
Eq.~(\ref{eqn:dgdq2}) can be rewritten as
\begin{equation}
    \label{eqn:dndq2}
    \frac{dn}{dq^2} = A N_{\mathrm{tag}} p^3 |f_+(q^2)|^2,
\end{equation}
where $A = \frac{1}{2} \frac{G_F^2|V_{cs}|^2}{24\pi^3} \tau_{D^+}$, and
the number of observed semileptonic signal events as a function of
$q^2$ is given by
\begin{equation}
    \label{eqn:dn-dq2}
    \frac{dn_{\mathrm{observed}}}{dq^2} = A N_{\mathrm{tag}}
    \left[
      p^3(q'^2) |f_+(q'^2)|^2 \eff(q'^2)
    \right]
    \otimes \sigma(q'^2, q^2),
\end{equation}
where $q'^2$ refers to the true value and $q^2$ refers to the measured
value; $p(q'^2)$ is the momentum of $\kl$ in the rest frame of the
parent $D$; $\eff(q'^2)$ is the detection efficiency and
$\sigma(q'^2, q^2)$ is the detector resolution. To account for
detector effects, we use the theoretical function convoluted
with a Gaussian detector resolution to describe the observed signal
curve.

\subsection{Form-factor parametrizations}
\label{subsec:ff para}
The goal of any particular parametrization $f_+(q^2)$ of the
semileptonic form factors is to provide an accurate, and physically
meaningful, expression of the strong dynamics in the decays. One
possible way to achieve this goal is to express the form factors in
terms of a dispersion relation. This approach of using
dispersion relations and dispersive bounds in the description of form
factors, has been well established in the literature. In general, the
dispersive representation is derived from the evaluation of the two point
function~\cite{boyd-PRL, boyd-PRD} and can be written as
\begin{equation}
    f_+(q^2) = \frac{f_+(0)}{(1-\alpha)}\frac{1}{1 - \frac{q^2}{m_{\mathrm{pole}}^2}} + \frac{1}{\pi}\int^{\infty}_{(m_D + m_P)^2}\frac{\mathrm{Im}f_+(t)}{t-q^2-i\eff}dt,
\end{equation}
where $m_D$ and $m_P$ are the masses of the $D$ meson and pseudoscalar meson respectively, while $m_{\mathrm{pole}}$ is the mass of the lowest-lying $c\bar q$ vector meson, with $c \to q$ the quark transition of the semileptonic decay. For the charm semileptonic decays we have $m_{\mathrm{pole}}$ = $m_{D_s^*}$ for $D \to K e \nu_e$ decays. The parameter $\alpha$ expresses the size of the vector meson pole contribution to $f_+(0)$. It is common to write the contribution from the continuum integral as a sum of effective poles
\begin{equation}
    \label{eqn:semilep}
    f_+(q^2) = \frac{f_+(0)}{(1-\alpha)}\frac{1}{1 - \frac{q^2}{m_{\mathrm{pole}}^2}} + \sum^{N}_{k=1}\frac{\rho_k}{1 - \frac{1}{\gamma_k}\frac{q^2}{m_{\mathrm{pole}}^2}},
\end{equation}
where $\rho_k$ and $\gamma_k$ are expansion parameters.

The simplest parametrization, known as the simple pole model, assumes that the sum in Eq.~(\ref{eqn:semilep}) is dominated by a single pole
\begin{equation}
    \label{eqn:simple-pole}
    f_+(q^2) = \frac{f_+(0)}{1 - \frac{q^2}{m_{\mathrm{pole}}^2}},
\end{equation}
where the value of $m_{\mathrm{pole}}$ is predicted to be $m_{D_s^*}$. In experiments, $m_{\mathrm{pole}}$ is left as a free fit parameter to improve the fit quality.

Another parametrization is known as the modified pole model, or Becirevic-Kaidelov (BK) parametrization~\cite{becirevic}. The idea is to add the first term in the effective pole expansion, while making simplifications such that the form factor can be determined with only two parameters: the intercept $f_+(0)$ and an additional shape parameter $\alpha$. The simplified one-term expansion is usually written in the form
\begin{equation}
    \label{eqn:modified-pole}
    f_+(q^2) = \frac{f_+(0)}{(1 - \frac{q^2}{m_{\mathrm{pole}}^2})(1 - \alpha\frac{q^2}{m_{\mathrm{pole}}^2})}.
\end{equation}

A third parametrization is known as the series expansion~\cite{becher}. Exploiting the analytic properties of $f_+(q^2)$, a transformation of variables is made that maps the cut in the $q^2$ plane onto a unit circle $|z| < 1$, where
\begin{equation}
    \label{eqn:t0}
    z(q^2, t_0) = \frac{\sqrt{t_+ - q^2} - \sqrt{t_+ - t_0}}{\sqrt{t_+ - q^2} + \sqrt{t_+ - t_0}},
\end{equation}
$t_{\pm} = (m_D \pm m_P)^2$, and $t_0$ is any real number less than $t_+$. This transformation amounts to expanding the form factor about $q^2 = t_0$, with the expanded form factor given by
\begin{equation}
    \label{eqn:series-expansion}
    f_+(q^2) = \frac{1}{P(q^2)\phi(q^2, t_0)}\sum_{k=0}^{\infty}{a_k(t_0)[z(q^2, t_0)]^k},
\end{equation}
where $a_k$ are real coefficients, $P(q^2) = z(q^2, M_{D_s^*}^2)$ for kaon final states, $P(q^2) = 1$ for pion final states, and $\phi(q^2, t_0)$ is any function that is analytic outside a cut in the complex $q^2$ plane that lies along the $x$-axis from $t_+$ to $\infty$. This expansion has improved convergence properties over Eq.~(\ref{eqn:semilep}) due to the smallness of $z$; for example, taking the traditional choice of
$t_0 = t_+(1 - (1 - t_-/t_+)^{1/2})$, which minimizes the maximum value of $z(q^2, t_0)$. Further, taking the standard choice of $\phi$:
\begin{equation}
    \label{eqn:series-expansion-phi}
    \begin{aligned}
    \phi(q^2, t_0) =& \sqrt{\frac{\pi m_c^2}{3}}\left(\frac{z(q^2, 0)}{-q^2}\right)^{5/2}\left(\frac{z(q^2, t_0)}{t_0 - q^2}\right)^{-1/2}\\
    &\times\left(\frac{z(q^2, t_-)}{t_- - q^2}\right)^{-3/4}\frac{t_+ - q^2}{(t_+ - t_0)^{1/4}},
    \end{aligned}
\end{equation}
where $m_c$ is the mass of charm quark, it can be shown that the sum over all $k$ of $a_k^2$ is of order unity.

In practical use of the series expansion form factor, one often takes $k = 1$ and $k = 2$ in Eq.~(\ref{eqn:series-expansion}), which gives following two forms of the form factor.
\bitm
    \item 2 par. series expansion of form factor is given by
    \begin{equation}
    \label{eqn:series-expansion-2}
    f_+(q^2) = \frac{1}{P(q^2)\phi(q^2, t_0)}a_0(t_0)\left(1 + r_1(t_0)[z(q^2, t_0)]\right).
    \end{equation}
    It can be rewritten as
    \begin{equation}
    \label{eqn:series-expansion-2a}
    \begin{aligned}
    f_+(q^2) =& \frac{1}{P(q^2)\phi(q^2, t_0)}\frac{f_+(0)P(0)\phi(0, t_0)}{1+r_1(t_0)z(0,t_0)}\\
    &\times\left(1 + r_1(t_0)[z(q^2, t_0)]\right),
    \end{aligned}
    \end{equation}
    where $r_1 = a_1 / a_0$.

    \item 3 par. series expansion of form factor is given by
    \begin{equation}
    \label{eqn:series-expansion-3}
    \begin{aligned}
    f_+(q^2) =& \frac{1}{P(q^2)\phi(q^2, t_0)}a_0(t_0)\\
    &\times\left(1 + r_1(t_0)[z(q^2, t_0)] + r_2(t_0)[z(q^2, t_0)]^2\right).
    \end{aligned}
    \end{equation}
    It can be rewritten as
    \begin{equation}
    \label{eqn:series-expansion-3a}
    \begin{aligned}
    f_+(q^2) =& \frac{1}{P(q^2)\phi(q^2, t_0)}\frac{f_+(0)P(0)\phi(0, t_0)}{1+r_1(t_0)z(0,t_0)+r_2(t_0)z^2(0,t_0)}\\
    &\times\left(1 + r_1(t_0)[z(q^2, t_0)] + r_2(t_0)[z(q^2, t_0)]^2\right),
    \end{aligned}
    \end{equation}
    where $r_1 = a_1 / a_0$, $r_2 = a_2 / a_0$.
\eitm

\subsection{\boldmath Determination of $f^{K}_+(0)|V_{cs}|$}
\begin{figure}[htbp]
    \centering
    \includegraphics[height=6cm]{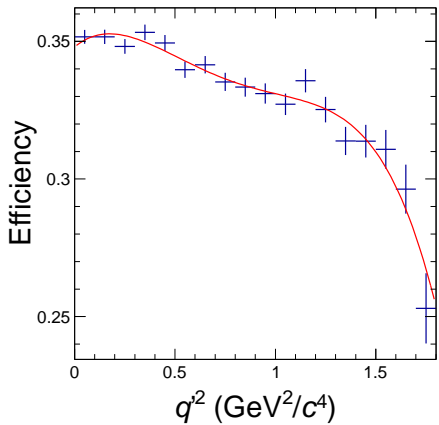}
    \caption{\label{fig:eff}Detection efficiency $\eff(q'^2)$ for
      signal events tagged by $D^{\pm} \to K^{\mp} \pi^{\pm}
      \pi^{\pm}$. The dots with error bars are the corrected signal MC
      efficiencies, and the curve is the fit result.}
\end{figure}

We perform simultaneous fits to the distributions of observed DT
candidates as a function of $q^2$ for the six ST modes to determine
$f^{K}_+(0)|V_{cs}|$. In the fits, we treat $D^+$ and $D^-$ DT
candidates together. The detection efficiency $\eff(q'^2)$ and
detector resolution $\sigma(q'^2, q^2)$ are obtained from the $\kl$
efficiency corrected signal MC simulations. For each ST mode,
$\eff(q'^2)$ is described by a fourth-order polynomial; the ($q^2 -
q'^2$) distribution is described by a Gaussian function. As an
example, Figure~\ref{fig:eff} shows the fits to $\eff(q'^2)$ for
signal events tagged by $D^{\pm} \to K^{\mp} \pi^{\pm} \pi^{\pm}$.

\begin{figure*}[htbp]
	\centering
    \includegraphics[height=3.5cm]{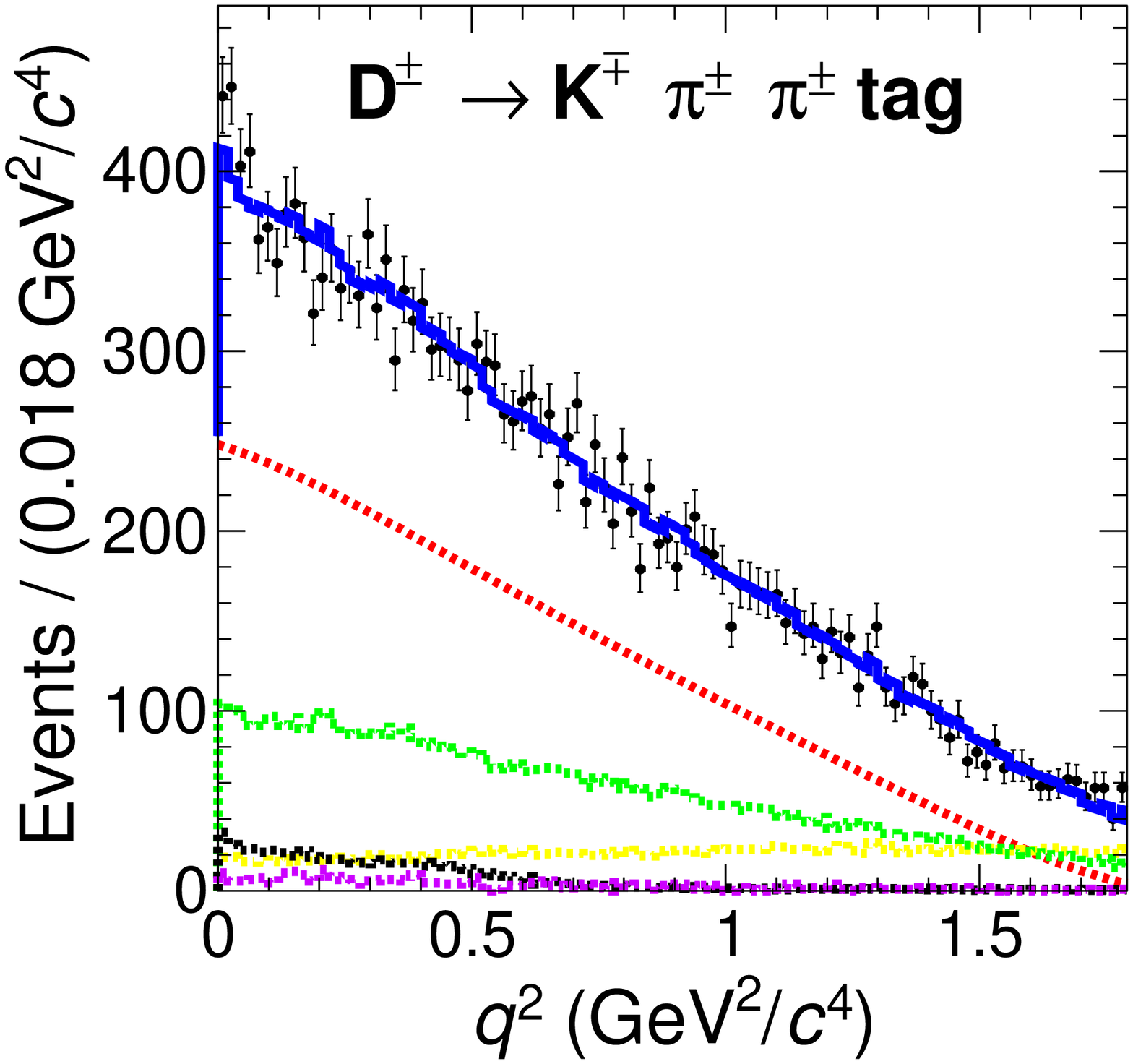}
    \includegraphics[height=3.5cm]{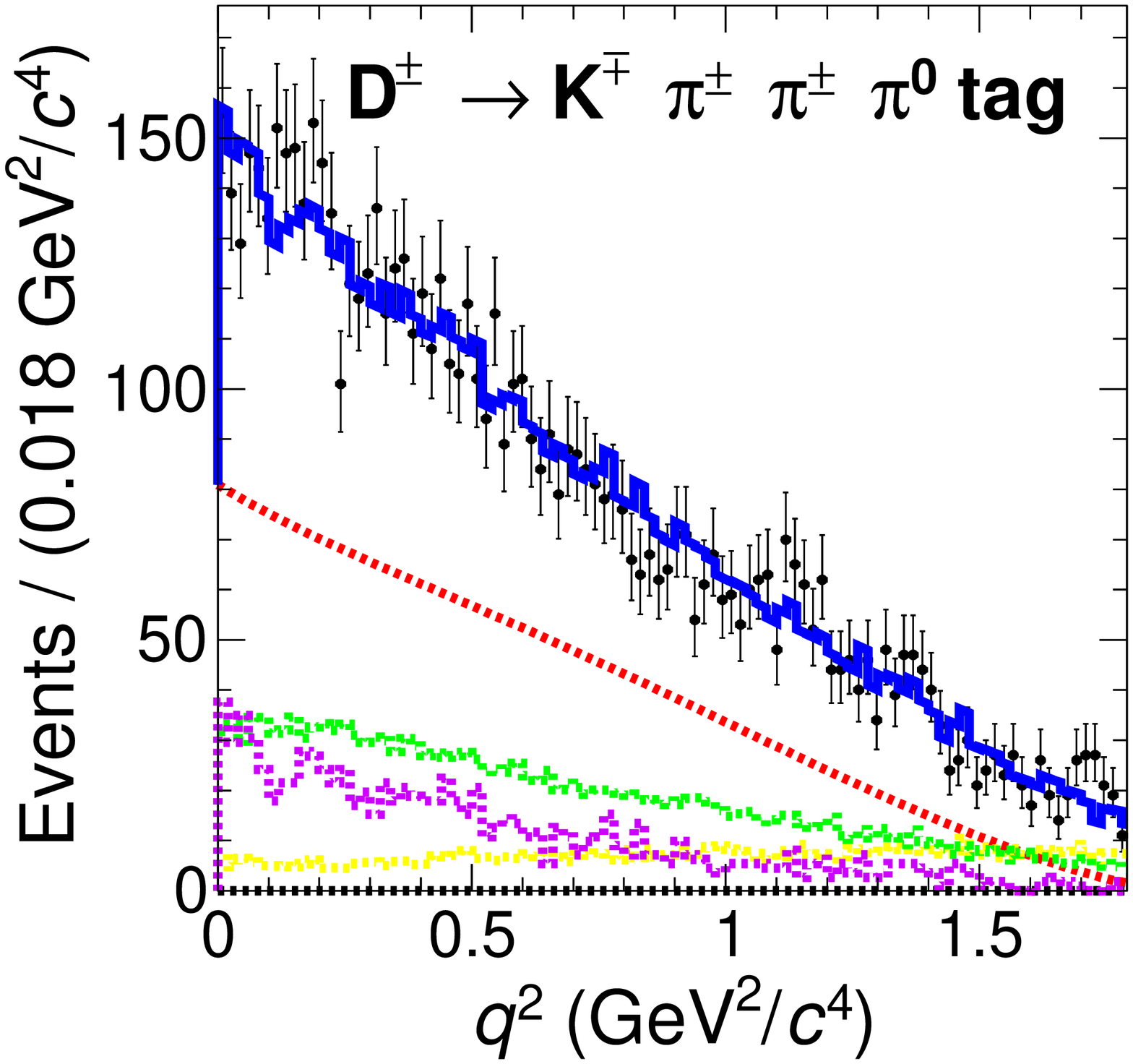}
    \includegraphics[height=3.5cm]{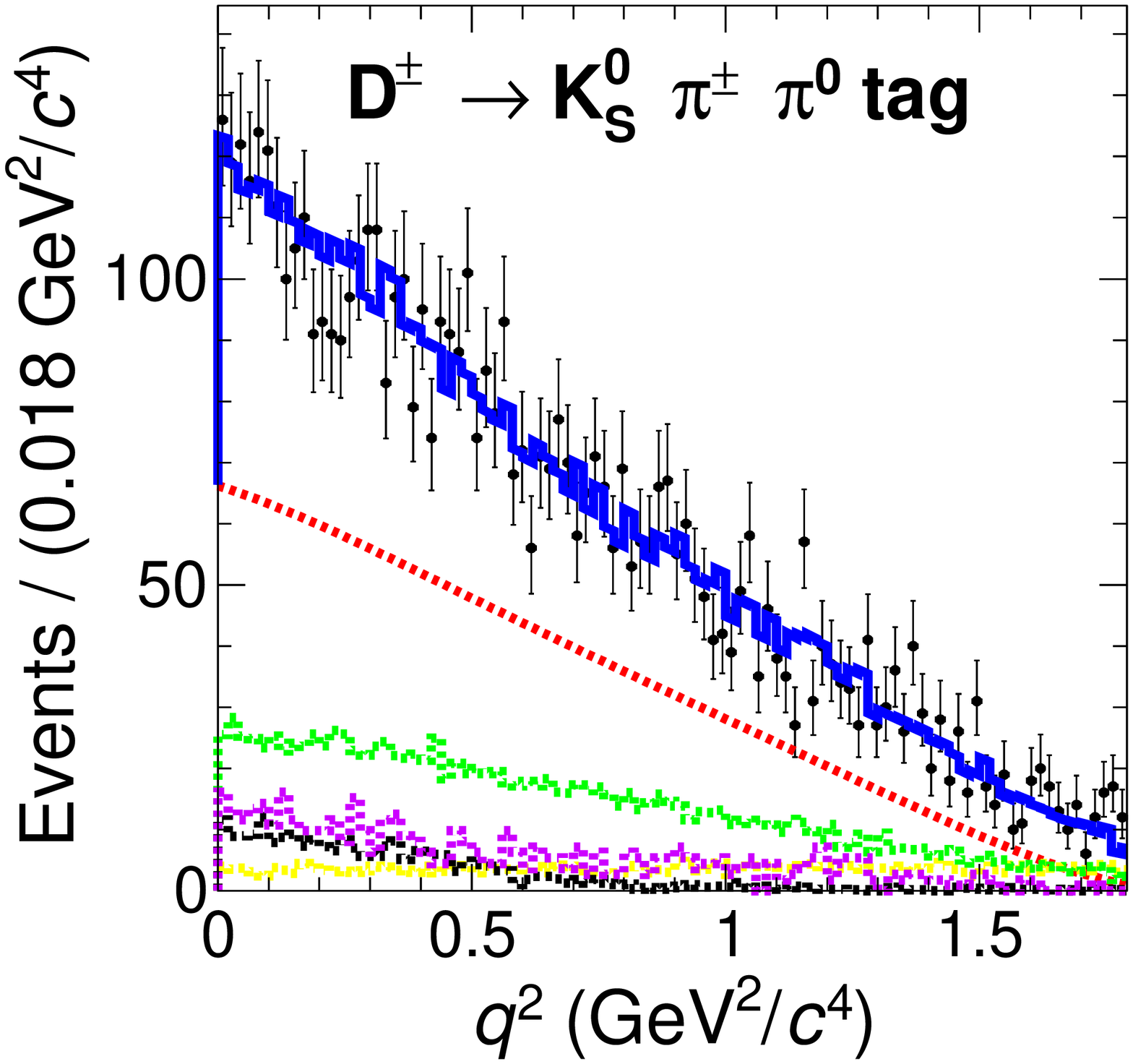}\\
    \includegraphics[height=3.5cm]{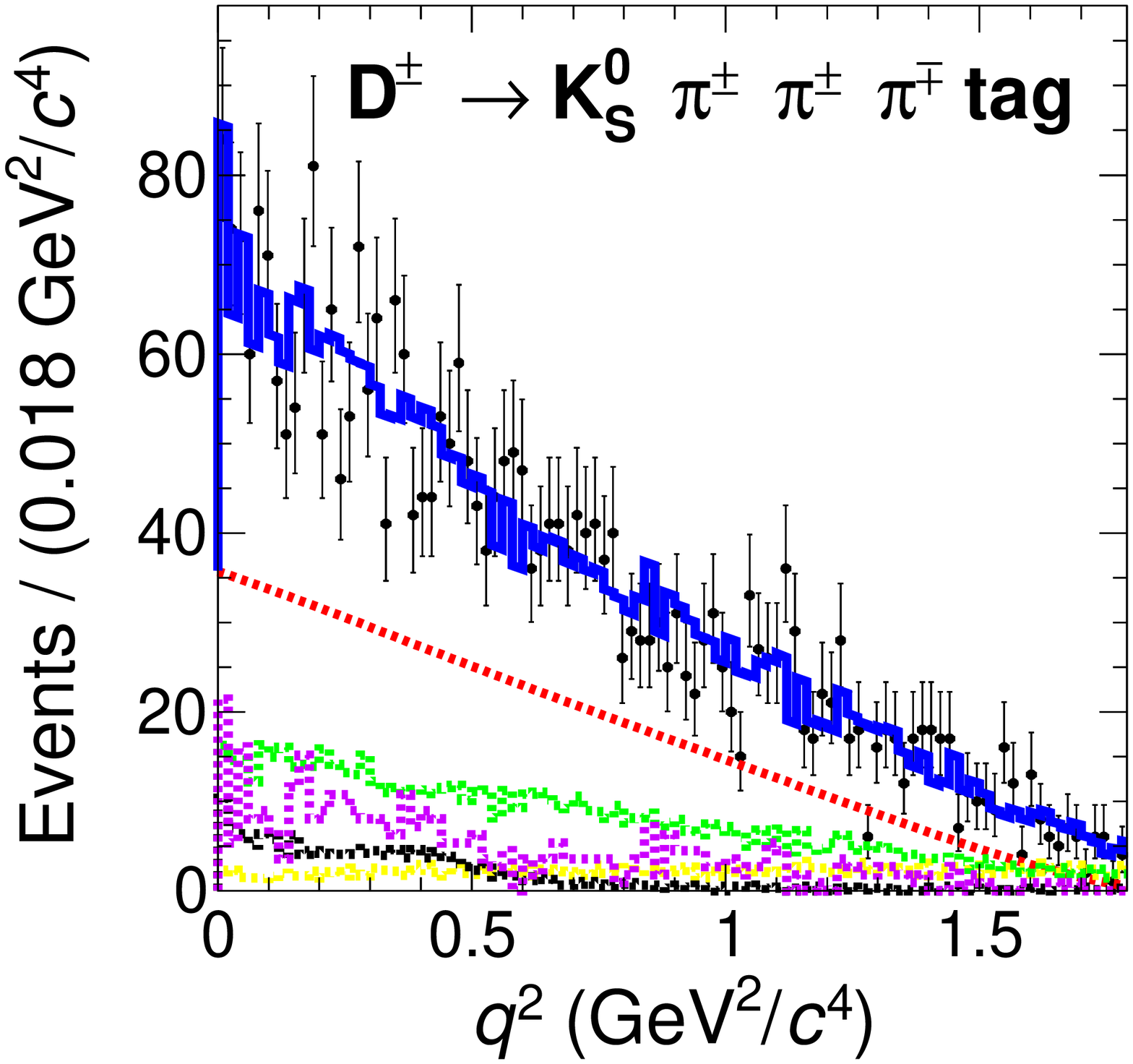}
    \includegraphics[height=3.5cm]{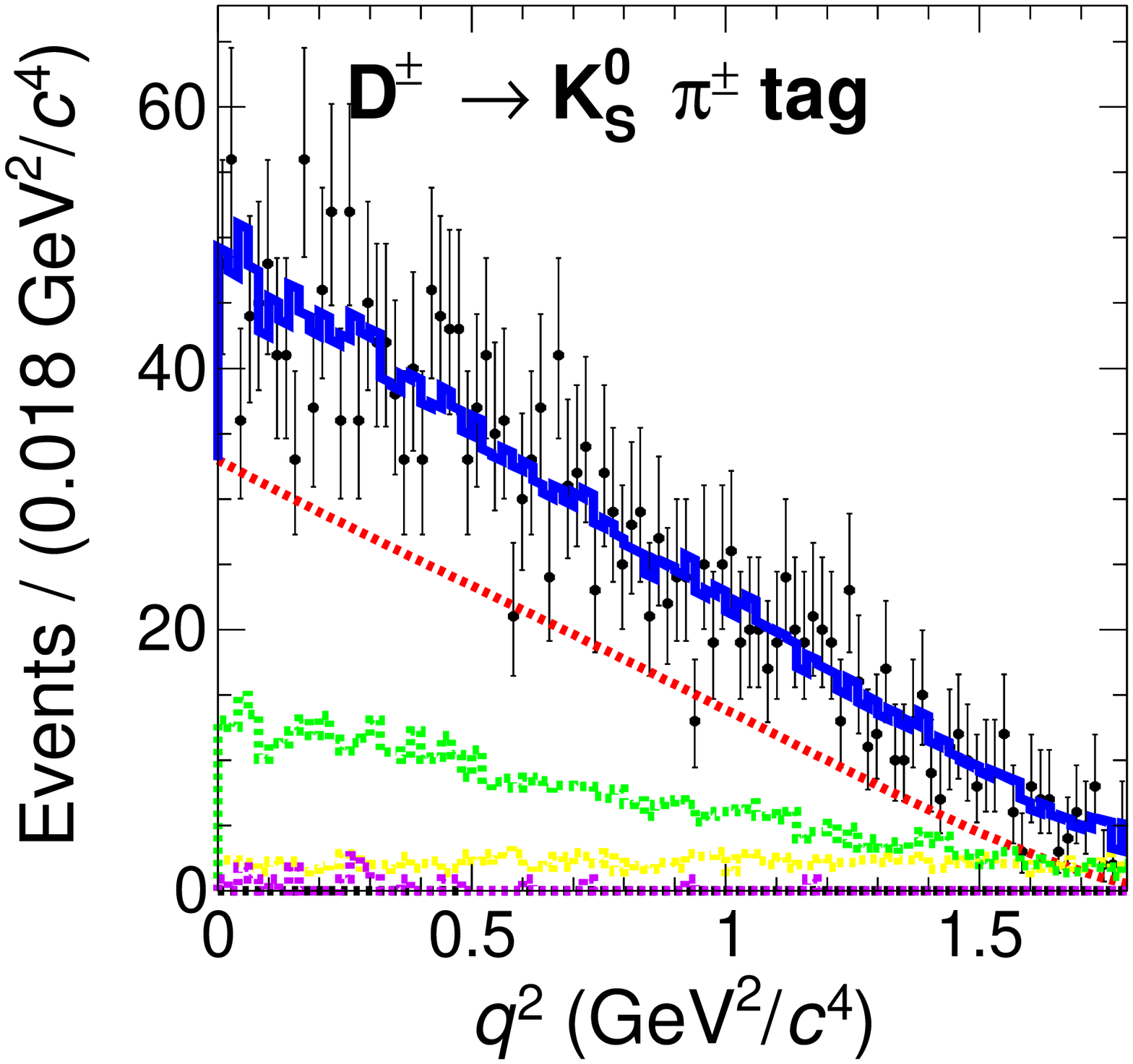}
    \includegraphics[height=3.5cm]{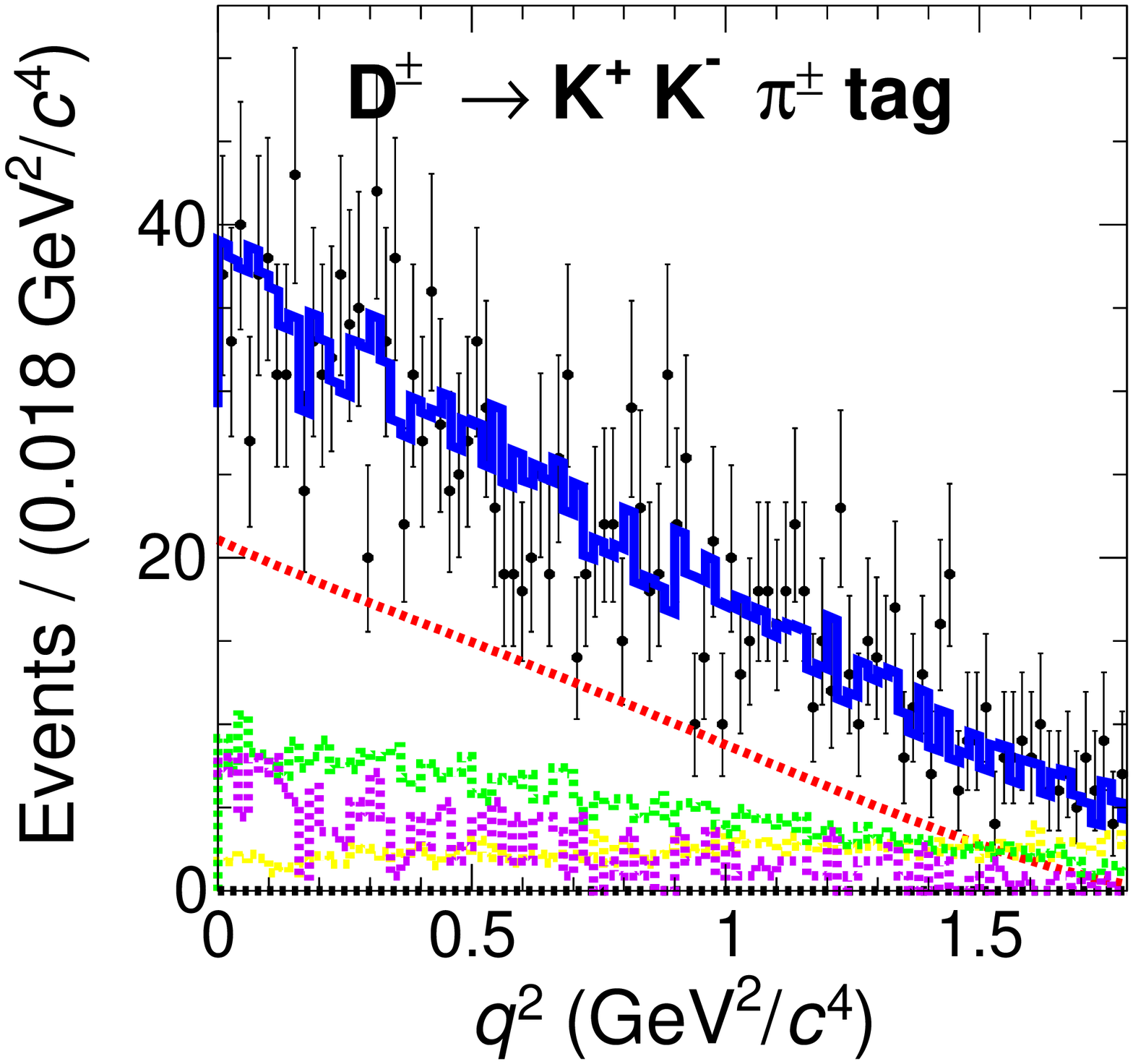}
	\caption{\label{fig:fit-q2-two}(Color online) Simultaneous
	fit to the numbers of DT candidates as a function of $q^2$
	with the two-parameter series expansion parametrization. The
	points are data and the curves are the fit to data. In each plot,
	the violet, yellow, green, and black curves refer to
	Bkg~\RN{1}, Bkg~\RN{2}, Bkg~\RN{3}+Bkg~\RN{4}, and Bkg~\RN{5},
	respectively. The red dashed curve shows the contribution of
	signal, and the blue solid curve shows the sum of background
	and signal.}
\end{figure*}

Simultaneous fits are made with one or two common parameters related
to the form-factor shape to the data for the simple pole model
($m_{\mathrm{pole}}$), the modified pole model ($\alpha$),
two-parameter series expansion ($r_1$) and three-parameter series
expansion ($r_1, r_2$). As an example, Figure~\ref{fig:fit-q2-two}
shows the simultaneous fit results using the two-parameter series
expansion model. The signal PDF is constructed in the form of
Eq.~(\ref{eqn:dn-dq2}). For the background shape, as mentioned in
Section~\ref{subsec:peakbkg}, the shape and the number of Bkg~\RN{1}
events are fixed according to the side-band region of the $\mbc$
distribution ($1.83 < \mbc < 1.85\GeV/c^2$) from data; for Bkgs from
\RN{2} to \RN{5}, the shape is determined from the $\kl$ efficiency
corrected generic MC samples. We also fix the relative proportion of
$N_{\mathrm{sig}}$, $N_{\mathrm{Bkg\,\RN{2}}}$ and
$N_{\mathrm{Bkg\,\RN{3}}}+N_{\mathrm{Bkg\,\RN{4}}}$ events, to the
result from the $\kl$ efficiency corrected generic MC. Here,
$N_{\mathrm{sig}}$, $N_{\mathrm{Bkg\,\RN{2}}}$,
$N_{\mathrm{Bkg\,\RN{3}}}$ and $N_{\mathrm{Bkg\,\RN{4}}}$ represent
the number of the signal, Bkg~\RN{2}, Bkg~\RN{3} and Bkg~\RN{4}
events, respectively.

The product $f^{K}_{+}(0)|V_{cs}|$ is obtained from
\begin{equation}
    \label{eqn:f+0}
    f^{K}_{+}(0)|V_{cs}| = \sqrt{\frac{48
    \pi^3}{G_F^2}\frac{N_{\mathrm{sig}}}{N_{\mathrm{tag}}\tau_{D^+}I}},
\end{equation}
where $I = \int \left[p^3(q'^2) |f_+(q'^2)|^2 \eff(q'^2)\right] \otimes
\sigma(q'^2, q^2) dq^2$.

Since the $q^2$ distribution of the signal events is smooth, the
form-factor fit is insensitive to the detector resolution. For each
tag mode, we use the full width at half maximum (FWHM) of the ($q^2 -
q'^2$) distribution to estimate $\sigma(q'^2, q^2)$ and obtain FWHM =
$0.0360\GeV^2/c^4$ and the corresponding resolution $\sigma =
\mathrm{FWHM}/2\sqrt{2\ln2} = 0.0153\GeV^2/c^4$. The distributions of
DT candidates as a function of $q^2$ are fit again by different models with
the detector resolution $\sigma = 0.0153\GeV^2/c^4$. Compared to
the previous results, the form-factor parameters and the signal yields
are almost unchanged. So the uncertainty of the detector resolution
can be ignored in the form-factor fit.

Systematic uncertainties of the form-factor parameters are more
sensitive to the distribution of backgrounds in this analysis. We use different side-band region of the $\mbc$
distribution ($1.835 < \mbc < 1.855\GeV/c^2$) and ISGW2 model to simulate the main possible semi-leptonic and semi-muonic backgrounds. We simultaneously fit the the distributions of observed DT candidates as a function $q^2$ again. The differences between the form-factor parameters obtained from the two determinations are taken as the systematic uncertainties of the form-factor parameters.

Systematic uncertainties associated with the product
$f^{K}_+(0)|V_{cs}|$ are one half of the systematic uncertainties in
the branching fraction measurements, presented in
Sec.~\ref{sec:brfraction}, combined in quadrature with the
uncertainties associated with $D^+$ lifetime ($0.67\%$)~\cite{pdg} and
the integration $I$, which are obtained by varying the form-factor
parameters by $\pm 1\sigma$.  The systematic uncertainties of
$f^{K}_+(0)|V_{cs}|$ are obtained for the simple pole model, modified
pole model, two-parameter series expansion and three-parameter series
expansion to be $1.4\%$, $1.5\%$, $1.5\%$, $1.2\%$, respectively.

The fit results are given in Table~\ref{table:formfactor}. As a
comparison, Table~\ref{table:formfactor} also lists the corresponding
form-factor results determined for $D^+ \to \ks e^+ \nu_e$ from
CLEO-$c$~\cite{besson}. Our results are consistent with those from
CLEO-$c$ within uncertainties except for three-parameter series
expansion model due to heavy backgrounds in this analysis. In general,
as long as the normalization and at least one shape parameter are
allowed to float, all models describe the data well. We choose the
two-parameter series fit to determine $f^{K}_+(0)$ and $|V_{cs}|$.

The BESIII experiment has recently reported the most precise value of
$f^{K}_+(0)|V_{cs}|$ using the two-parameter series expansion for $D^0 \to
K^- e^+ \nu_e$~\cite{rongg}.  It is in agreement with the results
reported here.

\begin{table*}[htbp]
    \centering
    \scriptsize
    \caption{\label{table:formfactor}Comparison of results of
    $f^{K}_+(0)|V_{cs}|$ and shape parameters ($m_{\mathrm{pole}}$,
    $\alpha$, $r_1$ and $r_2$) to previous corresponding results
    determined by $D^+ \to \ks e^+ \nu_e$ from
    CLEO-$c$~\cite{besson}. The first uncertainties are statistical,
    and the second are systematic.}
    \begin{tabular}{lccc}\hline\hline
        \multicolumn{4}{c}{Single pole model} \\
        Decay mode               & $f^{K}_+(0)|V_{cs}|$         & $m_{\mathrm{pole}}$ ($\GeV/c^2$) &        \\
        $D^+ \to \kl e^+ \nu_e$  & $0.729 \pm 0.006 \pm 0.010$  & $1.953 \pm 0.044 \pm 0.036$      &        \\
        $D^+ \to \ks e^+ \nu_e$  & $0.720 \pm 0.006 \pm 0.009$  & $1.95 \pm 0.03 \pm 0.01$         &        \\
        \hline
        \multicolumn{4}{c}{Modified pole model} \\
        Decay mode               & $f^{K}_+(0)|V_{cs}|$         & $\alpha$                         &        \\
        $D^+ \to \kl e^+ \nu_e$  & $0.727 \pm 0.006 \pm 0.011$  & $0.239 \pm 0.077 \pm 0.065$      &        \\
        $D^+ \to \ks e^+ \nu_e$  & $0.715 \pm 0.007 \pm 0.009$  & $0.28 \pm 0.06 \pm 0.02$         &        \\
        \hline
        \multicolumn{4}{c}{Two-parameter series expansion} \\
        Decay mode               & $f^{K}_+(0)|V_{cs}|$         & $r_1$                            &        \\
        $D^+ \to \kl e^+ \nu_e$  & $0.728 \pm 0.006 \pm 0.011$  & $-1.91 \pm 0.33 \pm 0.28$        &        \\
        $D^+ \to \ks e^+ \nu_e$  & $0.716 \pm 0.007 \pm 0.009$  & $-2.10 \pm 0.25 \pm 0.08$        &        \\
        \hline
        \multicolumn{4}{c}{Three-parameter series expansion} \\
        Decay mode               & $f^{K}_+(0)|V_{cs}|$         & $r_1$                            & $r_2$                   \\
        $D^+ \to \kl e^+ \nu_e$  & $0.737 \pm 0.006 \pm 0.009$  & $-2.23 \pm 0.42 \pm 0.53$        & $11.3 \pm 8.5 \pm 8.7$  \\
        $D^+ \to \ks e^+ \nu_e$  & $0.707 \pm 0.010 \pm 0.009$  & $-1.66 \pm 0.44 \pm 0.10$        & $-14 \pm 11 \pm 1$      \\
        \hline\hline
    \end{tabular}
\end{table*}

\subsection{Determination of $f^{K}_+(0)$ and $|V_{cs}|$}
Using the $f^{K}_+(0)|V_{cs}|$ value from the two-parameter series
expansion fit and $|V_{cs}| = 0.97343 \pm 0.00015$ from PDG fits
assuming CKM unitarity~\cite{pdg} or $f_+^K(0) = 0.747 \pm 0.019$ from
the unquenched LQCD calculation~\cite{lqcd} as input, we obtain
\begin{equation}
    \label{eqn:fk}
    f^{K}_+(0) = 0.748 \pm 0.007 \pm 0.012
\end{equation}
and
\begin{equation}
    \label{eqn:vcs}
    |V_{cs}| = 0.975 \pm 0.008 \pm 0.015 \pm 0.025,
\end{equation}
where the uncertainties are statistical, systematic, and external (in
Eq.~(\ref{eqn:vcs})).  For Eq.~(\ref{eqn:fk}), the external error is
negligible ($0.0002$) compared to our measurement. The measured
$f^{K}_+(0)$ is consistent with the one measured with $D^+ \to \ks e^+
\nu_e$ at CLEO-$c$~\cite{besson}; it is also in good agreement with
LQCD predictions, although the currently available LQCD results have
relatively large uncertainties. The measured $|V_{cs}|$ is in
agreement with that reported by the PDG.

\section{Summary}
\label{sec:summary}

In this paper we present the first measurement of the absolute
branching fraction $\mathcal{B}(D^+ \to \kl e^+ \nu_e) = (4.481 \pm
0.027(\mathrm{stat.}) \pm 0.103(\mathrm{sys.}))\%$, which is in
excellent agreement with $\mathcal{B}(D^+ \to \ks e^+ \nu_e)$ measured
by CLEO-$c$~\cite{besson}. The $CP$ asymmetry $A_{CP}^{D^+ \to \kl e^+
\nu_e} = (-0.59 \pm 0.60(\mathrm{stat.}) \pm 1.48(\mathrm{sys.}))\%$,
which agrees with theoretical prediction on $CP$ violation in $K^0$ system within the statistical error, is also
determined. By fitting the distributions of the observed DT events as
a function of $q^2$, $f^{K}_+(0)|V_{cs}|$ and the corresponding
parameters for three different theoretical form-factor models are
determined. Taking $f_K^+(0)|V_{cs}|$ from the two-parameter series
expansion parametrization, $f_K^+(0)|V_{cs}| = 0.728 \pm
0.006(\mathrm{stat.}) \pm 0.011(\mathrm{sys.})$ and using $|V_{cs}|$
from the SM constraint fit, we find $f_{+}^{K}(0) = 0.748 \pm
0.007(\mathrm{stat.}) \pm 0.012(\mathrm{sys.})$. By using an unquenched
LQCD prediction for $f_{+}^{K}(0)$, $|V_{cs}| = 0.975 \pm
0.008(\mathrm{stat.}) \pm 0.015(\mathrm{sys.}) \pm
0.025(\mathrm{LQCD})$.

\begin{acknowledgments}
The BESIII collaboration thanks the staff of BEPCII and the IHEP
computing center for their strong support. This work is supported in
part by National Key Basic Research Program of China under Contract
No. 2015CB856700; National Natural Science Foundation of China (NSFC)
under Contracts Nos. 11125525, 11235011, 11322544, 11335008, 11425524;
the Chinese Academy of Sciences (CAS) Large-Scale Scientific Facility
Program; the CAS Center for Excellence in Particle Physics (CCEPP);
the Collaborative Innovation Center for Particles and Interactions
(CICPI); Joint Large-Scale Scientific Facility Funds of the NSFC and
CAS under Contracts Nos. 11179007, U1232201, U1332201; CAS under
Contracts Nos. KJCX2-YW-N29, KJCX2-YW-N45; 100 Talents Program of CAS;
National 1000 Talents Program of China; INPAC and Shanghai Key
Laboratory for Particle Physics and Cosmology; German Research
Foundation DFG under Contract No. Collaborative Research Center
CRC-1044; Istituto Nazionale di Fisica Nucleare, Italy; Ministry of
Development of Turkey under Contract No. DPT2006K-120470; Russian
Foundation for Basic Research under Contract No. 14-07-91152; The
Swedish Resarch Council; U.S. Department of Energy under Contracts
Nos. DE-FG02-04ER41291, DE-FG02-05ER41374, DE-SC0012069, DESC0010118;
U.S. National Science Foundation; University of Groningen (RuG) and
the Helmholtzzentrum fuer Schwerionenforschung GmbH (GSI), Darmstadt;
WCU Program of National Research Foundation of Korea under Contract
No. R32-2008-000-10155-0. This work is also supported by the NSFC
under Contract Nos. 11275209, 11475107.
\end{acknowledgments}

\appendix
\section{Systematic uncertainty in $\kl$ reconstruction efficiency}
\label{app:kl-sys}

To determine the systematic uncertainty in the $\kl$ reconstruction
efficiency, we measure the $\kl$ efficiency in data and MC using a
partial reconstruction technique. We then determine the efficiency
difference between data and MC,
$\eff_{\mathrm{data}}/\eff_{\mathrm{MC}} - 1$, of the $\kl$
reconstruction efficiency, where $\eff_{\mathrm{MC}}$ is the
efficiency for MC and $\eff_{\mathrm{data}}$ is the efficiency for
data.

Based on 1.3~B $\jpsi$ events collected by BESIII detector in years
2009 and 2012, we use two control samples to measure $\kl$
reconstruction efficiency. One sample is $\jpsi \to K^{*}(892)^{\pm}
K^{\mp}$ with $K^*(892)^{\pm} \to \kl \pi^{\pm}$, and the other is
$\jpsi \to \phi \kl K^{\pm} \pi^{\mp}$. We reconstruct all
the particles in the event except the $\kl$ whose efficiency we wish
to measure. The number of $K^0$($\bar K^0$) is denoted by $N_1$. Then,
by applying $\kl$ selection requirements mentioned in
Sec.~\ref{subsec:signal}, we obtain the number of $K^0$($\bar K^0$)
denoted by $N_2$. Here, in order to select $\kl$ control samples with
low level of backgrounds, we perform the kinematic fit to select $\kl$
candidate with the minimal $\chi^2$ and require $\chi^2 < 100$.

$K^0$($\bar K^0$) reconstruction efficiency is calculated by $\eff = N_2 / N_1$. For data, $N_1$, $N_2$ are determined by fitting the missing mass squared distribution of $\kl$. Each fit included a signal line shape function which is determined from MC samples smeared with a Gaussian resolution, and the background shape is determined from MC samples as well. With respect to MC samples, $N_1$, $N_2$ are obtained from MC truth directly. The fits are performed in separate momentum bins. In each fit, $N_1$ ($N_2$) consists of the number of $\kl$ and $\ks$. The ratio of $\kl$ to $\ks$ is estimated from MC simulations. Due to the effect of the difference in nuclear interactions of $K^0$ and $\bar K^0$ mesons, we consider $K^0 \to \kl$ and $\bar K^0 \to \kl$ separately. We use the charge of kaon to tag $K^0$ or $\bar K^0$ in the control sample, which means if we find a $K^+$ in the process, the corresponding $\kl$ must be derived from $\bar K^0$.

\begin{figure}[htbp]
	\centering
    \includegraphics[width=0.25\textwidth]{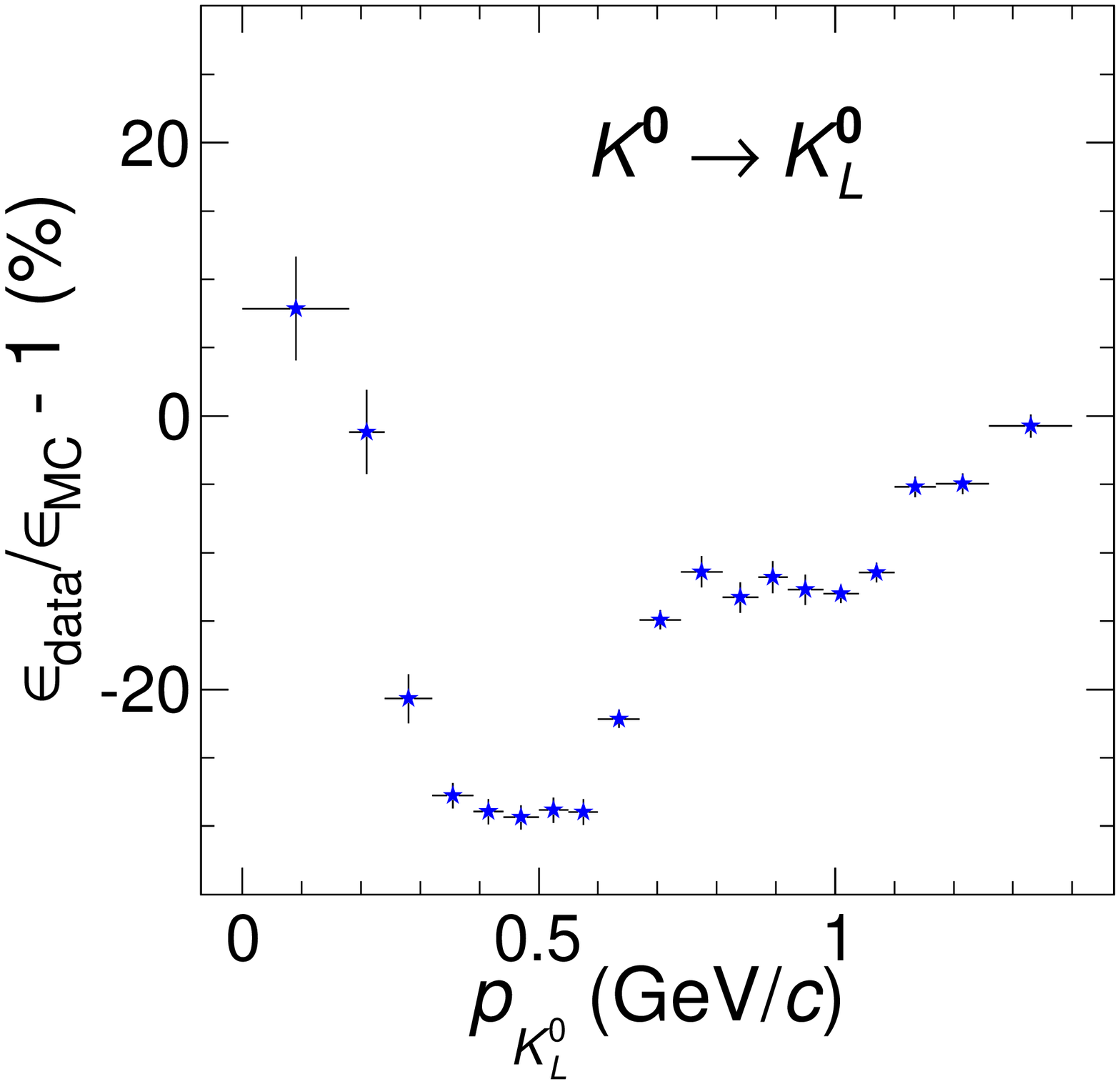}%
    \includegraphics[width=0.25\textwidth]{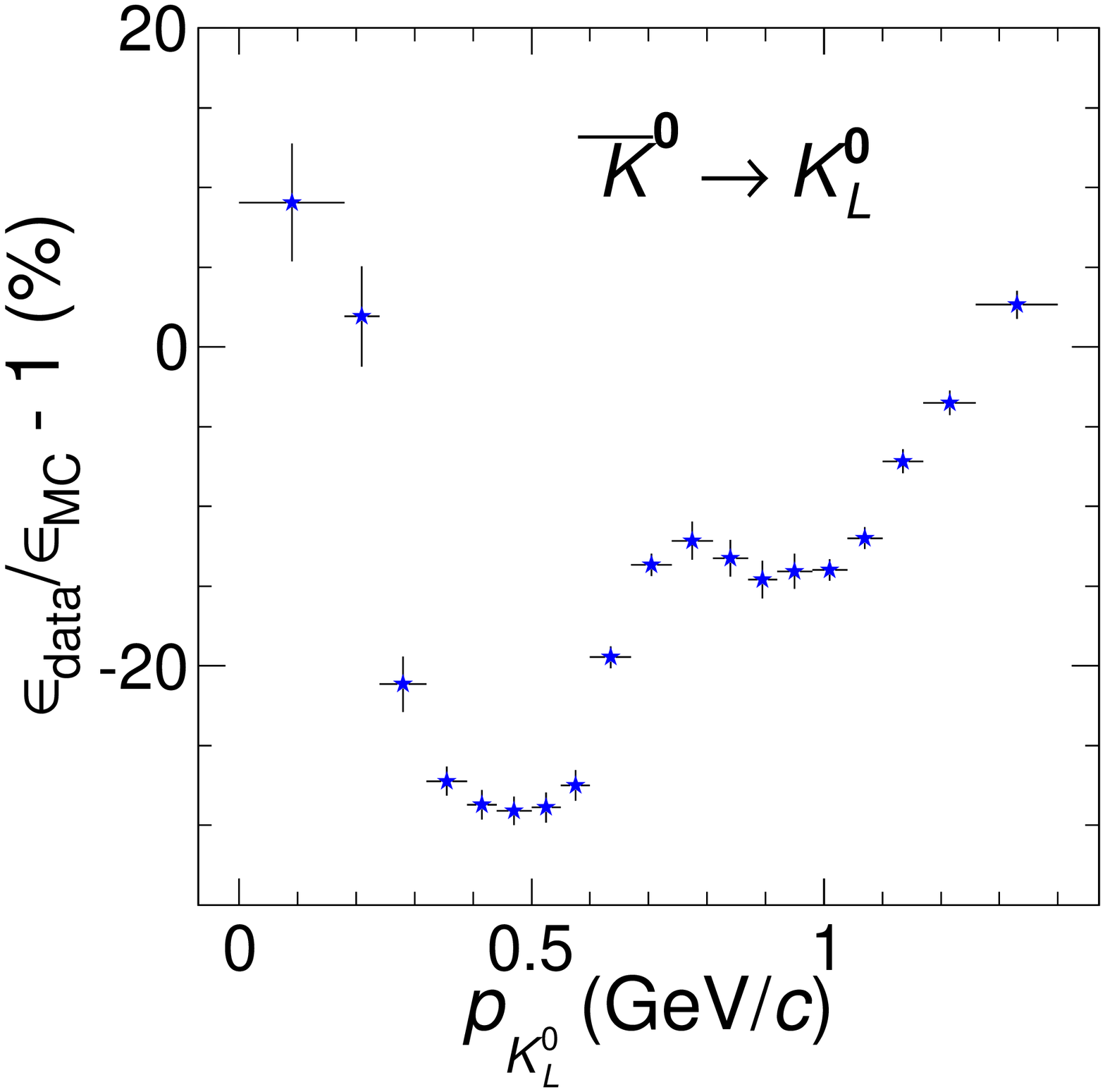}
	\caption{\label{fig:kl-eff-comparison}Distributions of $\kl$ reconstruction efficiency differences between data and MC for the processes of $K^0 \to \kl$ and $\bar K^0 \to \kl$.}
\end{figure}

Figure~\ref{fig:kl-eff-comparison} shows the distributions of $\kl$ reconstruction efficiency differences between data and MC in 19 momentum bins for the processes of $K^0 \to \kl$ and $\bar K^0 \to \kl$.

\begin{figure}[htbp]
	\centering
    \includegraphics[width=0.25\textwidth]{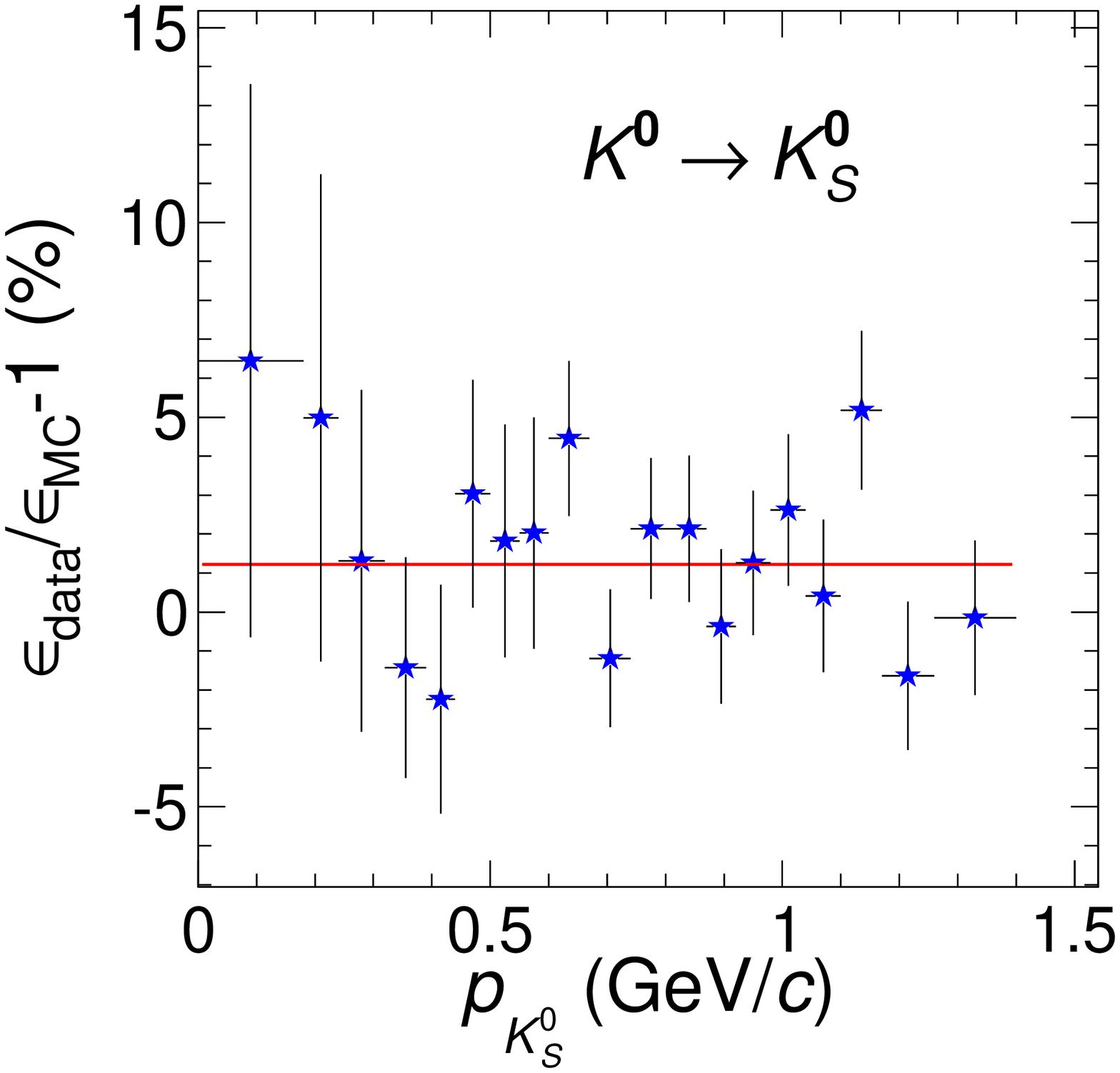}%
    \includegraphics[width=0.25\textwidth]{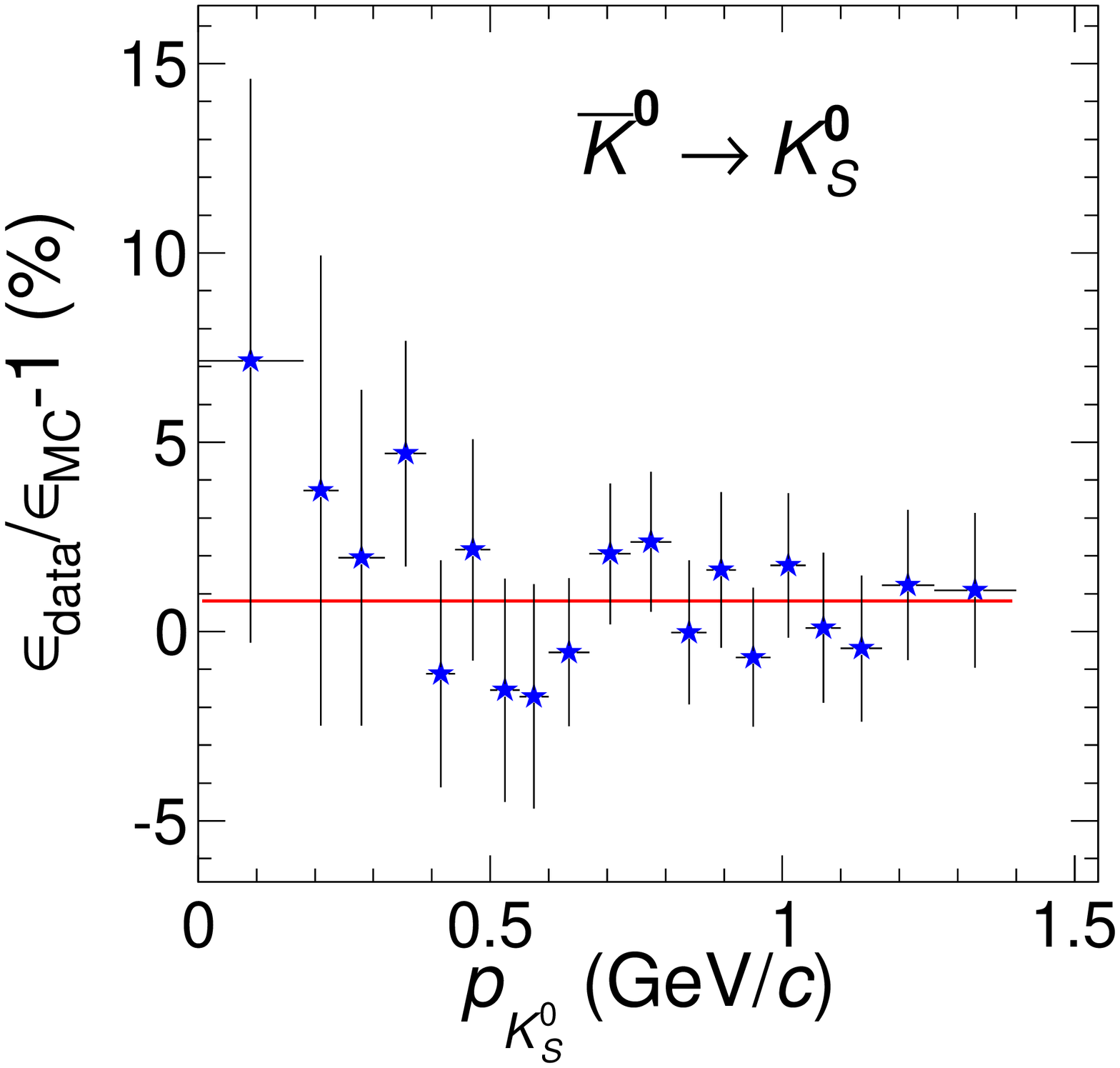}
	\caption{\label{fig:ks-sys}Distributions of $\ks$ reconstruction efficiency differences between data and MC for the processes of $K^0 \to \ks$ and $\bar K^0 \to \ks$. The red line is the fit to the points in the form of zero-order polynomial.}
\end{figure}

The probability of an inelastic interaction of a neutral kaon in the detector depends on the strangeness of the kaon at any point along its path, which is due to oscillations in kaon strangeness and different nuclear cross sections for $K^0$ and $\bar K^0$. Hence, the total efficiency to observe a final state $\kl$($\ks$) differs from that expected for either $K^0$ or $\bar K^0$. This effect is related to the coherent regeneration of neutral kaons~\cite{pais}. However, the detector-simulation program GEANT4 does not take into account this effect. The time-dependent $K^0$-$\bar K^0$ oscillations are thereby ignored in GEANT4. Considering the massive detector materials in the outer of the MDC, the TOF counter and the EMC, it results in an obvious discrepancy ($>$10$\%$) of $\kl$ shower-finding efficiency in the EMC between data and MC. On the other hand, we take the same method to study $\ks$ reconstruction efficiency difference between data and MC for the processes of $K^0 \to \ks$ and $\bar K^0 \to \ks$ by 224~M $\jpsi$ control sample, as shown in Fig.~\ref{fig:ks-sys}. We find that the $\ks$ reconstruction efficiency of data is a little higher than that of MC, which gives another hint of the absence of the coherent regeneration of neutral kaons by GEANT4.

\end{document}

%% file: authors_sep2015.tex
\author{
  \begin{small}
    \begin{center}
      M.~Ablikim$^{1}$, M.~N.~Achasov$^{9,f}$, X.~C.~Ai$^{1}$,
      O.~Albayrak$^{5}$, M.~Albrecht$^{4}$, D.~J.~Ambrose$^{44}$,
      A.~Amoroso$^{49A,49C}$, F.~F.~An$^{1}$, Q.~An$^{46,a}$,
      J.~Z.~Bai$^{1}$
% ...
      \\
      \vspace{0.2cm}
      (BESIII Collaboration)\\
      \vspace{0.2cm} {\it
        $^{1}$ Institute of High Energy Physics, Beijing 100049, People's Republic of China\\
% ...
$^{4}$ Bochum Ruhr-University, D-44780 Bochum, Germany\\
        $^{5}$ Carnegie Mellon University, Pittsburgh, Pennsylvania 15213, USA\\
% ...
$^{9}$ G.I. Budker Institute of Nuclear Physics SB RAS (BINP), Novosibirsk 630090, Russia\\
% ...
$^{44}$ University of Rochester, Rochester, New York 14627, USA\\
% ...
$^{46}$ University of Science and Technology of China, Hefei 230026, People's Republic of China\\
% ...
$^{49}$ (A)University of Turin, I-10125, Turin, Italy; (B)University of Eastern Piedmont, I-15121, Alessandria, Italy; (C)INFN, I-10125, Turin, Italy\\
% ...
$^{a}$ Also at State Key Laboratory of Particle Detection and Electronics, Beijing 100049, Hefei 230026, People's Republic of China\\
% ...
$^{f}$ Also at the Novosibirsk State University, Novosibirsk, 630090, Russia\\
% ...
      }
    \end{center}
    \vspace{0.4cm}
  \end{small}
}

\affiliation{}